%% file: MAIN_ModellingPaper_V16_rev_V3.tex
\documentclass[final,5p,times,twocolumn,sort, compress]{elsarticle} 
\usepackage{amsmath}
\usepackage{amssymb}
\usepackage{siunitx}
\graphicspath{{pics/}}						

\usepackage{ngerman}
\usepackage[hyphens]{url}
\usepackage{hyperref}  
\hypersetup{colorlinks = true, allcolors = black}
\usepackage{lineno}
\usepackage[11pt]{moresize}
\usepackage{graphicx}
\usepackage[dvipsnames]{xcolor}
\usepackage{booktabs}
\usepackage{multirow}
\usepackage{booktabs}
\usepackage{float}
\usepackage{appendix}
\usepackage{framed} 
\immediate\write18{%
	makeindex -s nomencl.ist -o \jobname.nls -t \jobname.nlg \jobname.nlo%
}

\usepackage{multicol} 

\usepackage{nomencl}

\makenomenclature
\renewcommand*\nompreamble{\begin{multicols}{3}}
	\renewcommand*\nompostamble{\end{multicols}}
\newlength{\nomgroupstartsep}
\usepackage{etoolbox}
\renewcommand\nomgroup[1]{%
	\itemsep\nomgroupstartsep%
	\item[\bfseries
	\ifstrequal{#1}{S}{Subscripts}{%
		\ifstrequal{#1}{H}{Superscripts}{%
			\ifstrequal{#1}{V}{Variables}{%
				\ifstrequal{#1}{A}{Abbreviations}{}}}}%
	]
	\itemsep\nomitemsep
}

\usepackage[tableposition=top]{caption}

\def\v#1{\mbox{\boldmath $\mathrm{ #1} $}}
\def\r#1{\mathrm{ #1 }}

\journal{Journal of Process Control}

\begin{document}
	
	\begin{frontmatter}
		
		
		
		\title{Control-oriented modeling of a $ \r{LiBr/H_2O} $ absorption heat pumping device \\ and experimental validation}
		
		
		\author[BEST,IRT]{Sandra Staudt}
		\author[BEST]{Viktor Unterberger}
		\author[BEST,IRT]{Markus~G{\"o}lles\corref{cor1}}
		\ead{markus.goelles@best-research.eu}
		\author[IWT]{Michael~Wernhart}
		\author[IWT]{Ren{é}~Rieberer}
		\author[BEST,IRT]{Martin~Horn}

		\cortext[cor1]{Corresponding author: Inffeldgasse 21/B, 8010 Graz, Austria}
		\address[BEST]{BEST - Bioenergy and Sustainable Technologies GmbH, Inffeldgasse 21/B, 8010 Graz, Austria}
		\address[IRT]{Institute of Automation and Control, Graz University of Technology, Inffeldgasse 21/B, 8010 Graz, Austria}
		\address[IWT]{Institute of Thermal Engineering, Graz University of Technology, Inffeldgasse 25/B, 8010 Graz, Austria}

		\begin{abstract}

			Absorption heat pumping devices (AHPDs, comprising absorption heat pumps and chillers) are devices that use thermal energy instead of electricity to generate heating and cooling, thereby facilitating the use of  waste heat and renewable energy sources such as solar or geothermal energy.
			 Despite this benefit, widespread use of AHPDs is still limited. One reason for this is partly unsatisfactory control performance under varying operating conditions, which can result in poor modulation and part load capability. A promising approach to tackle this issue is using dynamic, model-based control strategies, whose effectiveness, however,  strongly depend on the model being used. 
			This paper therefore focuses on the derivation of a viable dynamic model to be used for such model-based control strategies for AHPDs such as state feedback or model-predictive control. 
			The derived model is experimentally validated, showing good modeling accuracy. Its modeling accuracy is also compared to alternative model versions, that contain other heat transfer correlations, as a benchmark.
			Although the derived model is mathematically simple, it does have the structure of a nonlinear differential-algebraic system of equations. To obtain an even simpler model structure,  linearization at an operating point  is discussed to derive a model in linear state space representation.	The experimental validation shows that the linear model does have slightly worse steady-state accuracy,   but that the dynamic accuracy seems to be almost unaffected by the linearization.	The presented new  modeling approach is considered suitable to be used as a basis for the design of advanced, model-based control strategies, ultimately aiming to improve the modulation and part load capability of AHPDs.
			
			
			
			%

			
		\end{abstract}
		
		%
		
		
	\end{frontmatter}
	
	
	
		\nomenclature[v00]{$\Delta t_\r{d}$ }{dead time  [\si{\second }]}
	\nomenclature[v01]{$\epsilon$}{effectiveness {[-]}} 
	\nomenclature[v03]{$\phi_{\mathrm{sub}}$ }{ subcooling fraction [-]}
	\nomenclature[v04]{$\rho$}{density [\si{\kilogram \per \meter \cubed}]}
	\nomenclature[v05]{$\tau$ }{time constant  [\si{\second }]}
	\nomenclature[v06]{$\xi$  }{ mass fraction of LiBr in solution  [$\frac{\text{kg}_{\r{LiBr}}}{\text{kg}_{\r{solution}}}$]}
	
	\nomenclature[v07]{$c_{p}$ }{specific heat capacity  [\si{\joule \per \kilogram \per \kelvin }]}
	\nomenclature[v08]{$h$}{specific enthalpy [\si{\joule \per \kilogram }]}
		\nomenclature[v09]{$H$}{enthalpy [\si{\joule}]}
		
	\nomenclature[v13]{$m$  }{mass  [\si{\kilogram}]}
	\nomenclature[v14]{$\dot{m}$ }{mass flow rate  [\si{\kilogram \per \second}]}
	\nomenclature[v15]{$p$}{pressure [\si{\pascal}]}
	\nomenclature[v16]{$\dot{Q}$  }{heat flow rate [\si{\watt}]}
	\nomenclature[v18]{$t$ }{time  [\si{\second }]}
	\nomenclature[v19]{$T$  }{temperature [\si{\kelvin}]}

	\nomenclature[v21]{$\mathit{TTD}$  }{terminal temperature difference [\si{\kelvin}]}
			\nomenclature[v23]{$U$}{internal energy   [\si{\joule }]}

	\nomenclature[v32]{$\mathit{UA}$ }{$\mathit{UA}$ value  [\si{\watt \per \kelvin }]}
	\nomenclature[v33]{$\dot{V}$ }{volume flow rate [\si{\meter \cubed \per \second}]}

	\nomenclature[s01]{A}{absorber}
	\nomenclature[s02]{C}{condenser}
	\nomenclature[s03]{E}{evaporator}
	\nomenclature[s04]{G }{generator}
	\nomenclature[s05]{GRh }{gas room on high pressure side}
	\nomenclature[s06]{GRl }{gas room on low pressure side}
	\nomenclature[s07]{h}{ high}
	\nomenclature[s08]{$\mathrm{H_2O}$ }{water}
	\nomenclature[s09]{HX}{heat exchanger }
	
	\nomenclature[s11]{in }{inflowing}
	\nomenclature[s12]{l}{low}
	\nomenclature[s13]{LiBr}{lithium bromide}
	\nomenclature[s14]{meas}{measured }
	\nomenclature[s16]{out}{outflowing}
	\nomenclature[s17]{PSo}{poor solution }
	\nomenclature[s18]{rec}{recirculating}
	
	\nomenclature[s21]{Ref}{refrigerant}
	\nomenclature[s22]{ROP}{at reference operating point }
	\nomenclature[s23]{RSo}{rich solution }
	\nomenclature[s24]{sat}{saturated }
	\nomenclature[s25]{SHX}{solution heat exchanger}
		\nomenclature[s26]{ss}{steady state}
				\nomenclature[s27]{sumps}{in sumps}
	\nomenclature[s28]{W}{water in external circuits}


	\nomenclature[h]{l}{liquid }
	\nomenclature[h]{v}{vapor }

	\section{Introduction}
	\label{sec:intro}

	Heat pumping devices (HPDs, comprising heat pumps and chillers) are  devices that can transfer thermal energy from a low to a high temperature level. This requires an energetic input, either in the form of mechanical work (usually supplied by an electric motor) in the case of compression HPDs or in the form of high-temperature thermal energy in the case of absorption HPDs (AHPDs). 
	Therefore, AHPDs can use waste heat and renewable energy sources such as solar or geothermal energy instead of electricity to generate heating and cooling in a resource-efficient manner.

	Despite this benefit, AHPDs are still not a very common technology since operators are often put off by the increased complexity of AHPDs (e.g. more in- and output variables compared to compression HPS) and a lack of dynamic control strategies that allow good control performance over a wide operating range and that consider the coupled dynamics of different system variables   \cite{Goyal2019}. 	
	Current control strategies for AHPDs typically use SISO PI controllers or rely on simple ON/OFF operation, which may be sufficient for many current AHPD applications with rather constant operating conditions. However, they can reach their limits when AHPDs are integrated into modern, more complex energy systems where renewable, volatile energy sources play an increasingly important role. 
Here, one way to further advance AHPD control is to design dynamic model-based control strategies for AHPDs, e.g. \cite{Goyal2019a},  (such as state feedback or model-predictive control)  to extend their operating range.	
	This, however, first requires a suitable control-oriented, dynamic AHPD model - preferably in the form of a state-space model - which is the focus of this paper. The modeling approaches discussed here refer to AHPDs with the working fluid lithium bromide/water ($\r{LiBr/H_2O}$).
	 In the following, first an overview of available models from  literature is given, followed by the contribution this paper aims to make and its structure. 

	\subsection{Available models for $ {LiBr/H_2O} $ AHPDs}
	\label{subsec:available_models}

The literature on modeling of $ \r{LiBr/H_2O} $ AHPDs contains several reasonable approaches, including some that can be used as a basis for a control-oriented AHPD model. An overview shall be given subsequently:
Available models for AHPDs can first of all be divided into black-box models on the one hand and physically based models on the other hand.
In the former case, the authors of \cite{Lazrak2016} developed a dynamic artificial neural network model and successfully showed selected experimental validation results. In \cite{Chow2002} a neural network approach is used to model steady-state correlations between desired cooling capacity, selected disturbance variables and the corresponding necessary driving energy. Such black-box approaches have the disadvantage that they are not easily scalable though and thus have to be re-trained with a large number of (measurement) data to derive a robust model for a new machine, even if the AHPD design is similar and only the size differs from the reference machine.
Physically-based models, on the other hand, are easier to scale. 
They are also significantly more prevalent in the literature on AHPD modeling and can roughly be subdivided into three groups:

First, there are steady-state models, e.g.  \cite{Furukawa1987, Florides2003, Kim2008, PuigArnavat2010, Nia2015, Albers2018,  Rathod2019, Hosseinirad2019, Albers2021, Prieto2022},  which can be very useful in the dimensioning and design process, for feedforward-control of AHPD and for system-level control of energy systems with AHPD as shown in \cite{Rathod2019, Hosseinirad2019, Albers2021} but are not viable for dynamic model-based control methods since they do not capture the AHPD's dynamics. 

Second, there are dynamic, discretized models, i.e. the components of the AHPD are spatially discretized in the model, e.g. \cite{Fu2006, Bittanti2010,Vinther2015, Wernhart2020}. These are excellent models for very detailed simulation studies but are usually too complex and computationally expensive to be used for model-based control due to the high number of state variables and therefore the high model order.  In \cite{Vinther2015}, however, the authors discuss the possibility to use the presented high-order, nonlinear model, which was implemented in Dymola\textsuperscript{\textregistered} \cite{DassaultSystemes}, as a basis for a simpler, linear model by employing built-in functions of Dymola\textsuperscript{\textregistered} and Matlab\textsuperscript{\textregistered} \cite{MathWorks2021}  to first linearize the model and then reduce the model order, which is a reasonable approach to derive a control-oriented model in case a complex one is already available in Dymola\textsuperscript{\textregistered}. 

Third, there are simpler dynamic models based on so-called lumped-component approaches, e.g. \cite{Jeong1998, Kohlenbach2008,  Evola2013, Marc2015,  Sabbagh2018,Ochoa2016,Zinet2012,delaCalle2016, Castro2020}, where  individual components are modeled as lumped nodes with concentrated properties like temperature, density etc., which allows for a significantly lower model order compared to the discretized models mentioned before. Also combinations of lumped-component and discretized models, e.g. \cite{Misenheimer2017, Xu2016, Shin2009}, or  lumped-component and black-box models, e.g. \cite{Borg2012}, exist.
Within the group of lumped-component models the model complexity can roughly be assessed on the one hand by the used correlations for heat transfer and on the other hand by the number of state variables. Models in \cite{Jeong1998,Kohlenbach2008, Evola2013, Marc2015, Sabbagh2018} use constant heat transfer coefficients while \cite{Ochoa2016, delaCalle2016,Zinet2012, Castro2020} use more complex $ Nusselt $ correlations. 
The model order ranges from app. 12 \cite{Jeong1998} to 27 \cite{delaCalle2016} state variables, depending on which mass and energy stores are considered in the model. 
In principle, these lumped-component models (\cite{Jeong1998, Kohlenbach2008, Zinet2012, Evola2013, Marc2015, delaCalle2016, Ochoa2016, Sabbagh2018}) can be a good basis for control-oriented AHPD models. However, these models from literature are either not or not fully validated \cite{Jeong1998,Ochoa2016, Sabbagh2018, Zinet2012, delaCalle2016} so that it is not clear which model accuracy can be expected, or they do not describe dynamic effects of all manipulable input variables \cite{ Jeong1998, Kohlenbach2008, Evola2013, Marc2015, Ochoa2016}. In particular, one of the input variables, the flow rate of the so-called rich solution,  is either assumed to be constant \cite{ Jeong1998, Kohlenbach2008, Evola2013, Marc2015} or not considered in the validation process \cite{Sabbagh2018, Jeong1998, Ochoa2016, delaCalle2016, Zinet2012, Castro2020}. 
But since this is an input variable that shall be used as manipulated variable for model-based control, it is important that a control-oriented model can describe its dynamic effects.

	\subsection{Research contribution}

	The aim of this paper is to discuss control-oriented modeling approaches for $ \r{LiBr/H_2O} $ AHPDs that describe the main dynamics of the AHPD's output variables, being the outlet temperatures and heat flow rates in the hydraulic circuits, for variations of all relevant input variables, and compare it to measurement data from an AHPD at a laboratory test bench as well as to alternative model versions as a benchmark.
	The main research contributions can be summarized as: 
	\begin{enumerate}
		\item Derivation of a control-oriented AHPD model, taking into account varying operating conditions, especially variations of the rich solution flow rate
		\item Investigation of the effect of linearization at an operating point on modeling accuracy
		\item Extensive experimental validation for an AHPD with a nominal cooling capacity of \SI{15}{\kilo \watt} and comparison to alternative model versions as a benchmark.
	\end{enumerate}

	\subsection{Structure of paper}
	
	\autoref{sec:process_description} gives a process description for $\r{LiBr/H_2O}$ AHPDs in general and presents the specific AHPD that is used as a basis for the present work.  \autoref{sec:Sim_model} describes the new AHPD model and its linearization at a reference operating point. In \autoref{sec:benchmark_models}  two benchmark models are described which are used for comparison in the validation section \autoref{sec:validation} where all presented models are compared to each other and to measurement data to analyze the models' steady-state and dynamic accuracy. Lastly, a conclusion and outlook is given in \autoref{sec:conclusion}.

	\begin{table*}[h!]
		\begin{framed}
			\printnomenclature
		\end{framed}
	\end{table*}


	\section{Process description}
	\label{sec:process_description}

\begin{figure*}[h!]
		\centering
		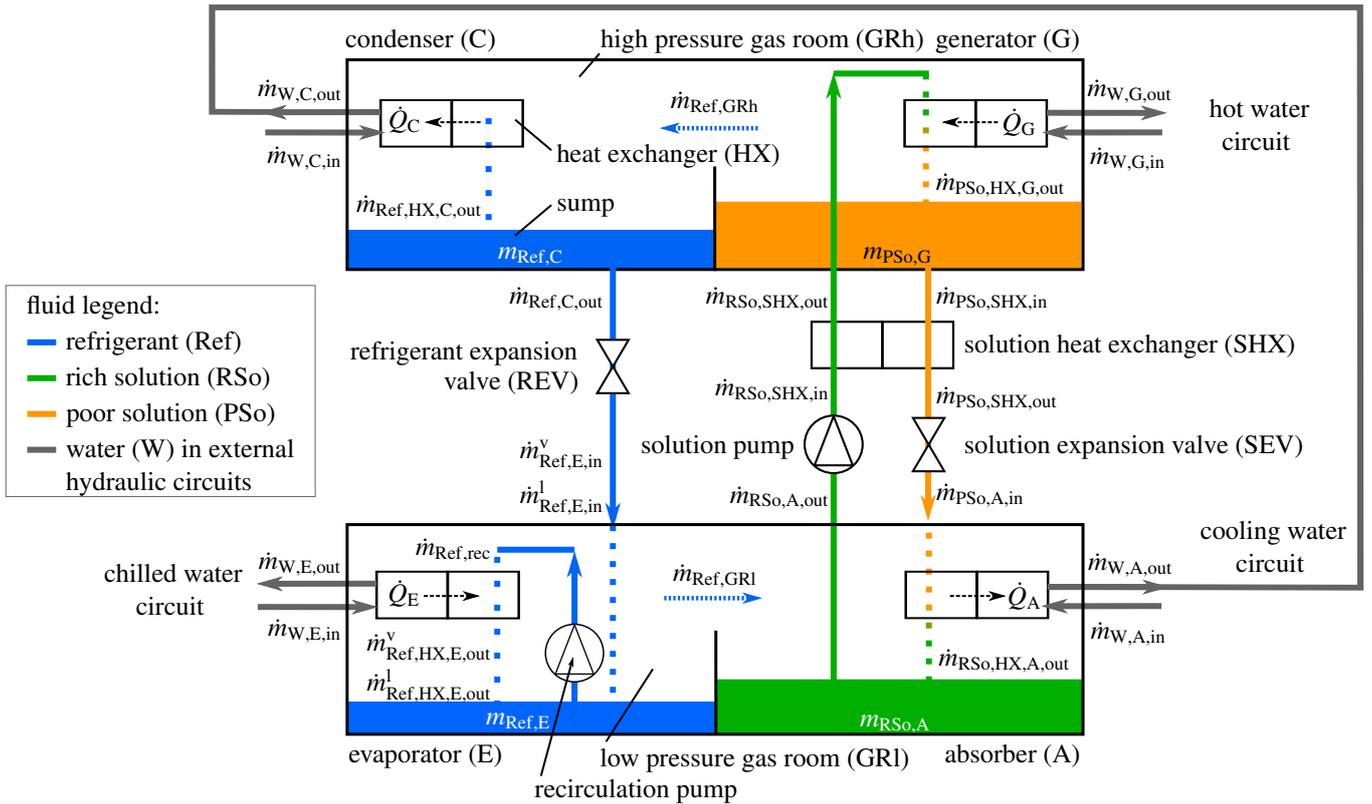
		\caption{Scheme of investigated $\r{LiBr/H_2O}$ AHPD and relevant mass and energy flows ($\dot{m}_{\r{subscript}}$) and stores (${m}_{\r{subscript}}$)}
		\label{fig:AHPD_Process}%
	\end{figure*}

	The models discussed in this paper are intended for $ \r{LiBr/H_2O} $ AHPDs  as depicted in  
	\autoref{fig:AHPD_Process}. The schematic layout in 	\autoref{fig:AHPD_Process}  gives an overview over relevant fluid streams (indicated by $ \dot{m}_{\r{subscript}} $, e.g. $  \dot{m}_{\r{PSo,SHX,in}}$) and fluid stores (indicated by $ m_{\r{subscript}} $, e.g. $ m_{\r{PSo,G}} $). For easier readability the following notation will be used throughout this paper: The two superscripts  l and v for liquid and vaporous fluid are only added when the corresponding fluid stream  consists of a liquid and a vaporous phase. In all other cases, when the fluid's state of aggregation is evident, the superscript will be omitted. 
	
	The most relevant components are the four main components generator, condenser, evaporator and absorber, each consisting of a heat exchanger (HX) and a fluid reservoir (sump),  the solution heat exchanger (SHX), the solution pump, the refrigerant recirculation pump, the solution expansion valve (SEV) and the refrigerant expansion valve (REV), where the SEV and REV can either be valves or simply pipes with reduced cross-sections.
	 Inside the AHPD, there are two interlinked circuits - a refrigerant circuit, containing $\r{H_2O} $ as refrigerant, and a solution circuit, containing $ \r{LiBr/H_2O} $, a mixture of $\r{H_2O} $ as refrigerant and LiBr as solvent. In addition, three external hydraulic circuits are connected to the AHPD for heat rejection (cooling water circuit) and heat supply (hot and chilled water circuit).
	The process has seven adjustable input variables - the inlet temperatures and  flow rates in the three hydraulic circuits, as well as the  flow rate of the rich solution.
	  One or more of the hydraulic circuits' outlet temperatures and/or a heat flow rate are usually the controlled variables of interest,  
	in detail depending on the specific application.
	
	The process can be described as follows. Starting at the generator, solution rich in refrigerant (rich solution) entering the generator ($ \dot{m }_{\r{RSo,SHX,out}} $) is heated up by the heat flow $ \dot{Q}_\r{G} $ and part of the absorbed refrigerant desorbs so that the solution leaving the HX ($ \dot{m }_{\r{PSo,HX,G,out}} $)  contains less refrigerant (poor solution). The desorbed refrigerant vapor ($ \dot{m }_{\r{Ref,GRh}} $) accumulates in the room between generator and condenser (high pressure gas room) and is condensed in the condenser ($ \dot{m }_{\r{Ref,HX,C,out}} $), thereby releasing the heat flow $ \dot{Q}_\r{C} $, and then flows into the condenser sump ($ {m }_{\r{Ref,C}} $). From here, it flows through the refrigerant expansion valve, where it expands to a lower pressure level, partly evaporates and cools down. The liquid part ($ \dot{m }^\r{l}_{\r{Ref,E,in}} $) of the now cold refrigerant flows into the sump of the evaporator ($ {m }_{\r{Ref,E}} $). The recirculation pump pumps refrigerant  over the evaporator HX ($ \dot{m }_{\r{Ref,E,rec}} $), where it partly evaporates by means of the heat flow $ \dot{Q}_\r{E} $. The liquid refrigerant leaving the evaporator HX ($ \dot{m }^\r{l}_{\r{Ref,E,HX,out}} $) flows back into the sump while the evaporated refrigerant ($ \dot{m }^\r{v}_{\r{Ref,E,HX,out}} $) and the vaporous refrigerant entering the evaporator ($ \dot{m }^\r{v}_{\r{Ref,E,in}} $) accumulate in the low pressure gas room. 
	In the absorber, poor solution flows over the HX ($ \dot{m}_{\r{PSo,A,in}} $), where it is cooled down by the heat flow $ \dot{Q}_\r{A} $ and absorbs this vaporous refrigerant ($ \dot{m}_{\r{Ref,GRl}} $).  The now rich solution ($ \dot{m}_{\r{RSo,A,HX,out}} $) flows into the absorber sump ($ m_{\r{RSo,A}} $).
	The solution pump then conveys the rich solution from the absorber ($ \dot{m}_{\r{RSo,A,out}} $) to the SHX ($ \dot{m}_{\r{RSo,SHX,in}} $) and into the generator ($ \dot{m}_{\r{RSo,SHX,out}} $),
	and raises its pressure to the high pressure level. In the SHX, heat is transferred from the poor solution to the rich solution, enhancing system efficiency. The poor solution flows from the generator sump ($ {m }_{\r{PSo,G}} $) through the SHX and the solution expansion valve, where it expands back to low pressure level, and finally into the absorber.   A detailed description of the working principle and other system designs can be taken from \cite{Herold2016}.
	When the AHPD is used as a chiller, the heat flow $ \dot{Q}_{\r{E}} $ is utilized and $ \dot{Q}_{\r{A}} $ and $ \dot{Q}_{\r{C}} $ have to be recooled by e.g. a cooling tower. When the AHPD is used as a heat pump, the heat flows $ \dot{Q}_{\r{A}} $ and $ \dot{Q}_{\r{C}} $ are utilized and $ \dot{Q}_{\r{E}} $ can be supplied by low temperature energy sources like e.g. waste or ambient heat. In either case the heat flow $ \dot{Q}_{\r{G}} $ has to be supplied by some high-temperature heat source.

	The proposed models in this paper are developed on the basis of and validated for the $\r{LiBr/H_2O} $ AHPD type \textit{WEGRACAL \textsuperscript{\textregistered} Maral 1} of the company EAW with a nominal cooling capacity of app. \SI{15}{\kilo \watt}  \cite{EAW2017} and  open plate heat exchangers in the generator, condenser, evaporator and absorber.

	\section{Control-oriented AHPD model}
	\label{sec:Sim_model}
	
	In this section the modeling of $\r{LiBr/H_2O} $ AHPDs for control purposes will be discussed. 
	First, the basic modeling approach is explained in \autoref{subsec:methodology}, followed by a discussion of relevant mass and energy stores in \autoref{subsec:rel_storage}. 
 Then, in \autoref{subsec:gen_con} to  \autoref{subsec:pumps}  the modeling of the AHPD's individual components is explained which are ultimately combined to form the complete AHPD model in \autoref{subsec:nonlin_complete}.
	
		For easier readability the following notation will be used from now on: When two variables with the same sub- and superscript are multiplied, they are joined in a square bracket with the common sub- and superscripts, so that e.g. $ \dot{m}^\r{l}_{\r{Ref,HX,E,out}} h^\r{l}_{\r{Ref,HX,E,out}}  $ becomes
	$ [\dot{m}h]^\r{l}_{\r{Ref,HX,E,out}}  $.

	\subsection{Basic modeling approach}
	\label{subsec:methodology}
%
	
	For the modeling in this paper lumped-component approaches are used, i.e.,  individual components are not modeled as spatially discretized components but considered as lumped nodes with concentrated properties, which allows for a rather simple model with few state variables. 
	Only those state variables, i.e., mass and energy stores,  that result in a dynamic effect in the relevant time range are considered, which will be discussed in \autoref{subsec:rel_storage}.
	The dynamic effects of these relevant mass and energy stores are described with transient mass and energy balances, i.e., ordinary first-order differential equations. In addition, algebraic equations are necessary to complete the model. These algebraic equations describe, e.g., heat transfer in the heat exchangers or correlations between physical quantities like, e.g., saturation pressure and temperature or enthalpy, temperature and mass fraction. 
	Due to the strong couplings between these algebraic equations and their generally non-linear character, they cannot easily be given as an explicit function of the input and state variables only but will be given in implicit form instead. 
	
	Moreover, the following modeling assumptions are used:
	\begin{itemize}
				\setlength\itemsep{-0.4em}
\item The refrigerant is pure water.
\item The refrigerant and solution leaving the generator, condenser and evaporator HXs are saturated, except the vaporous refrigerant leaving the generator which is assumed to have the same temperature as the entering rich solution. 
\item The expansion in the REV and SEV is isenthalpic.
\item All liquids are incompressible fluids. 
\item Heat losses are neglected. 
\item The power of the pumps is negligible.
	\end{itemize}
	

		\subsection{Discussion of relevant mass and energy stores}
	\label{subsec:rel_storage}

	For a simple model 	it is advisable to  assess which mass and energy stores result in  dynamic effects
	that occur on a time scale relevant for the latter control problem. 
	These include first-order delay effects, caused by mass or energy storage in a vessel,  and dead time effects, caused by the plug-flow of a fluid through a pipe or HX. 
	For this, dynamic effects with time constants $ \tau $ of app. \SIrange{5}{500}{\second} and dead times $ \Delta t_\r{d}  >$ \SI{5}{\second}  shall be considered relevant. 
So far, in the literature there does not seem to be a consistent approach to the choice of dynamics, which may be related to the fact that most models have been used for simulation  rather than for control purposes,  so that model order and complexity were less significant.
	Therefore, the individual mass and energy stores, are discussed below, based on the AHPD from \autoref{sec:process_description}:
	\begin{itemize}
		\setlength\itemsep{-0.4em}
		\item The room filled with vaporous refrigerant (high and low pressure gas room) contains a very small amount of refrigerant  and results in a first-order delay effect with time constants $\tau < \SI{1}{\s}  $. It is therefore modeled by means of algebraic equations in the form of steady-state mass and energy balances.
		\item Similarly, the low energy storage capability of the five HXs' metal plates in combination with relatively high heat transfer rates result in first-order delay effects with negligible time constants $\tau < \SI{5}{\s}  $ for the HX's metal plates. 
		\item The fluid in the main components' sumps have relevant dynamic effects with $ \SI{5}{\s}< \tau < \SI{500}{\s}$. They are approximated as first-order delay elements and modeled by first-order mass and energy balances. 
		\item The mass of refrigerant and solution at the surface of the main components' HXs is difficult to assess due to unknown void fractions of vaporous refrigerant in the two-phase liquids, but assumed to be sufficiently small in comparison to the mass in the sumps to neglect their first-order delay and dead time effects.
		\item The dead time effect of the water in the main components' HXs strongly depends on the operating conditions in the hydraulic circuits, so that for low volume flow rates  $\Delta t_\r{d}$ can reach \SI{5}{\s}. Additionally, the temperature sensors that measure in- and outlet temperatures are usually not mounted directly at the in- and outlet of the HXs but further away (due to constructural reasons), which results in additional dead times (up to \SI{20}{\s} for the investigated AHPD). For a more manageable control-oriented model these dead time effects are modeled in such a way that the main components' HXs are approximated as lumped nodes without dead time and instead the dead time effects are added to the hydraulic circuits by means of two volume-flow-dependent  dead time elements at the in- and outlet of each hydraulic circuit. These  are not considered part of the AHPD model itself, but are discussed with the test bench setup in \autoref{subsec:test bench} (\autoref{eq:deadtime_in} and \autoref{eq:deadtime_out}).
		\item The solution in the SHX and in the pipes between absorber and generator has relevant dynamic effects with volume-flow-dependent dead times of up to app. \SI{60}{\s} on both the poor and the rich solution side. These dead times occur ``in the middle'' of the AHPD as opposed to at the in- or outlet as discussed above and investigations showed that they can be approximated as first-order delay elements for a more manageable control-oriented model and still yield good dynamic modeling accuracy for the outlet temperatures and heat flow rates in the three hydraulic circuits.  
\item Similarly, the refrigerant in the pipe between condenser and evaporator would lead to relevant dead times. However, since this pipe basically connects two sumps, this dead time effect is neglected and instead the refrigerant is modeled as part of the consecutive evaporator sump, which in return is modeled as first-order delay element.  The same applies to the refrigerant in the pipes of the recirculation stream in the evaporator. Details on this simplification will be discussed in \autoref{subsubsec:REV}.
		\item The metal of the AHPD's shell has a high mass and energy storage capability but the heat transfer rate between fluid and shell is considerably lower than in the HXs. Its dynamic effect is therefore considered too slow to be considered in a control-oriented model.	Similarly, also heat losses to the ambient are neglected since they change with changing shell and ambient temperature, both being too slow to be considered. The effect of this simplification can easily be compensated by adding integral action to the model-based controller to be designed. 
	\end{itemize}

	\subsection{Generator and condenser}
	\label{subsec:gen_con}
	
	Generator and condenser each consist of a plate heat exchanger and a fluid sump and are housed in the same shell, whose gas room is filled with refrigerant vapor at high pressure (high pressure gas room). This subsystem's inlet streams are the rich solution flow from the SHX, the cooling water flow coming from the absorber and the hot water flow.
	
	\subsubsection{Heat exchangers}

	The incoming rich solution flows over the generator HX, is heated up by the incoming hot water flow and part of the absorbed refrigerant is desorbed. 
A very common way to model heat transfer in lumped-component AHPD models, e.g.  \cite{Jeong1998,Kohlenbach2008, Evola2013, Marc2015, Sabbagh2018, Ochoa2016}, is to use the HX's 
so-called $ \mathit{UA}$-value, which is the product of the HX's overall heat transfer coefficient $ U $ and  its heat transfer area $ A $. This HX parameter can be used to describe the heat transfer rate in dependence of the temperature difference between primary and secondary side of a HX. Usually, in thermal engineering the so-called logarithmic mean temperature difference,  is used for that - a modeling approach that can, however, result in singular behavior \cite{Herold2016} (e.g., when the two temperature differences at the HX's ends are equal) and is mathematically quite complex. Therefore, instead, many AHPD models make use of polynomial approximations of the logarithmic mean temperature difference, e.g. \cite{ Sabbagh2018}, or a linear approximation, i.e., the arithmetic mean temperature difference,  e.g. \cite{Kohlenbach2008,  Ochoa2016}, which is also used for the modeling in this paper. This simplification is deliberately accepted due to the mentioned possibility of singular behavior of the logarithmic mean temperature difference. 
	$ \dot{Q}_{\r{G}} $ is therefore calculated with the generator HX's $ \mathit{UA}$-value $\mathit{UA} _\r{G}$ and  the (arithmetic) mean  temperature difference $\overline{\Delta T}_{\r{HX,G}}$ as
	\begin{equation}
		\label{eq:Qdot_G}
		\dot{Q}_\r{G}=\mathit{UA} _\r{G} \overline{\Delta T}_{\r{HX,G}}
	 \end{equation} 
	\begin{equation}
		\label{eq:dT_G}
		\overline{\Delta T}_{\r{HX,G}}=\frac{T_{\r{W,G,in}}+T_{\r{W,G,out}}}{2}-\frac{T_{\r{RSo,SHX,out}}+T_{\r{PSo,HX,G,out}}}{2}
	 \end{equation} 
	consisting of the in- and outlet temperatures  of the solution, $ T_{\r{RSo,SHX,out}} $ and $ T_{\r{PSo,HX,G,out}} $, and of the hot water, $ T_{\r{W,G,in}} $ and $ T_{\r{W,G,out}} $. 
	Lumped-component models from  literature, see e.g. \cite{Jeong1998,Kohlenbach2008, Evola2013, Marc2015, Sabbagh2018}, often use constant $\mathit{UA}  $-values which, however, cannot capture the influence of varying mass flow rates in the HX. Instead, for the modeling in this paper, $ \mathit{UA} _\r{G} $ is described by 
	\begin{equation}
		\label{eq:UA_G}
		\mathit{UA} _\r{G}=   \r{K_{G1} }+ \r{K_{G2}} \dot{m}_{\r{W,G}} + \r{K_{G3}} \dot{m}_{\r{RSo,SHX,out}} + \r{ K_{G4}} \dot{m}_{\r{Ref,GRh}} 
	 \end{equation} 
	where $ \mathit{UA} _\r{G} $  increases with an increasing mass flow rate of hot water $ \dot{m}_{\r{W,G}} $, rich solution $\dot{m}_{\r{RSo,SHX,out}} $ and desorbed refrigerant $  \dot{m}_{\r{Ref,GRh}} $. 
	The parameters $ \r{K_{G1} }  $ to $  \r{K_{G4} } $ can be found empirically from measurement data or by means of more complex modeling approaches, like $ Nusselt $-correlations (see e.g. \cite{Ochoa2016, Zinet2012, delaCalle2016} for shell and tube heat exchangers; to the authors' knowledge there are no validated $ Nusselt $-correlations for two-phase heat transfer in open plate heat exchangers for $\r{LiBr/H_2O} $ yet.). 
	The parameters in this paper for the investigated AHPD were derived from measurement data at different operating points and are listed in \autoref{sec:param_empirical}.


	The heat flow $ \dot{Q}_\r{G} $ causes the hot water to cool down on the one hand and the solution to heat up and release part of the absorbed refrigerant on the other hand - a process that is of course subject to mass and energy conservation. 
	Since no energy storage is considered in the HX (see \autoref{subsec:rel_storage}) it is described by the following steady-state mass and energy balances: The mass flow of poor solution and refrigerant leaving the generator HX, $ \dot{m}_{\r{PSo,HX,G,out}} $ and $ \dot{m}_{\r{Ref,GRh}} $ must equal the entering mass flow of rich solution $ \dot{m}_{\r{RSo,SHX,out}} $:
\begin{equation}
	\label{eq:m_bal_HX_G}
0  	\left( =\frac{\r{d}m_{\r{So,HX,G}}}{\r{d}t} \right)= \dot{m}_{\r{RSo,SHX,out}}-\dot{m}_{\r{PSo,HX,G,out}}- \dot{m}_{\r{Ref,GRh}}
 \end{equation} 
Also the amount of LiBr leaving and entering the HX, $ [\dot{m} \xi]_{\r{PSo,HX,G,out}} $ and $  [\dot{m} \xi]_{\r{RSo,SHX,out}} $, must be equal:
\begin{equation}
	\label{eq:LiBr_bal_HX_G}
	0\left( =\frac{\r{d}m_{\r{LiBr,HX,G}}}{\r{d}t} \right)= [\dot{m} \xi]_{\r{RSo,SHX,out}} - [\dot{m} \xi]_{\r{PSo,HX,G,out}}
 \end{equation} 

Similarly, the energy balances on either side (hot water side and solution side) are as follows. Note that for the energy balances the stored enthalpy $ H $ instead of the internal energy $ U $ is considered since all liquids in this model are treated as incompressible fluids for which the difference between enthalpy and internal energy is negligible \cite{Moran2006}. 
\begin{equation}
		\label{eq:E_bal_HX_G_Sol}
\begin{split}
		0\left( = \frac{\r{d}H_{\r{So,HX,G}}}{\r{d}t} \right) = &[\dot{m} h]_{\r{RSo,SHX,out}}-[\dot{m} h]_{\r{PSo,HX,G,out}} \\
		& - [\dot{m} h]_{\r{Ref,GRh}} + \dot{Q}_\r{G}
\end{split}
 \end{equation} 
\begin{equation}
	\label{eq:E_bal_HX_G_W}
	0 \left( = \frac{\r{d}H_{\r{W,HX,G}}}{\r{d}t} \right)=\dot{m}_{\r{W,G}} (h_{\r{W,G,in}}-h_{\r{W,G,out}}) - \dot{Q}_\r{G}
 \end{equation} 
with the mass flow rate and specific enthalpy of the entering rich solution $[\dot{m} h]_{\r{RSo,SHX,out}}$, the leaving poor solution, $ [\dot{m} h]_{\r{PSo,HX,G,out}}$, the desorbed refrigerant $[\dot{m} h]_{\r{Ref,GRh}}  $ and of the entering and leaving hot water $\dot{m}_{\r{W,G}} (h_{\r{W,G,in}}-h_{\r{W,G,out}})$. 
Depending on whether volume or mass flow rates are used as input variables,
the  property functions below for density of $ \r{LiBr/H_2O} $ 	$ \rho_{\r{LiBr/H_2O}} $ and of liquid water $ \rho^{\r{l}}_{\r{H_2O}} $  can be used to calculated the mass flow rates  $\dot{m} _{\r{RSo,SHX,out}} $  and  $ \dot{m} _{\r{W,G}}  $ from corresponding volume flow rates if necessary.
\begin{equation}
	\label{eq:rho_LiBrH2O}
	\rho_{\r{LiBr/H_2O}}=\r{R_1} - \r{R_2} \xi+ \r{R_3} \xi^2 + \r{R_4} T 
 \end{equation} 
\begin{equation}
	\label{eq:rho_H2O}
	\rho^{\r{l}}_{\r{H_2O}}=\r{F_1} + \r{F_2} T- \r{F_3}  T^ 2
 \end{equation} 
 All  property  functions in this paper are simplified functions based on more complex  functions from  \cite{Yuan2005}  for $ \r{LiBr/H_2O} $ and from \cite{Wagner2002} for water. Their parameters are listed in \autoref{sec:param_substance_prop}.

The correlation between the specific enthalpies in \autoref{eq:E_bal_HX_G_Sol} and  \autoref{eq:E_bal_HX_G_W}, and the fluids' temperatures and 
mass fractions is  described by the following property functions for the solution, $h_{\r{LiBr/H_2O}}(T,\xi)$, the vaporous refrigerant, $h^\r{v}_{\r{H_2O}}(T)$, and the (liquid) hot water, $ h^\r{l}_{\r{H_2O}}(T) $:
\begin{equation}
	\label{eq:h_LiBrH2O}
	h_{\r{LiBr/H_2O}}(T,\xi)=-\r{A_1} - \r{A_2} \xi+\r{A_3} \xi^2 + \r{A_4}T - \r{A_5} \xi T 
 \end{equation} 
\begin{equation}
	\label{eq:h_v_H2O}
	h^\r{v}_{\r{H_2O}}(T)=\r{D_1} + \r{D_2}  T
 \end{equation} 
\begin{equation}
	\label{eq:h_l_H2O}
	h^\r{l}_{\r{H_2O}}(T)=-\r{C_1} + \r{C_2}  T
 \end{equation}


For the temperatures of the desorbed refrigerant, $ T_{\r{Ref,GRh}} $, and of the poor solution $ T_{\r{PSo,HX,G,out}} $ the following simplifying assumptions can be made: 
First, it is assumed that the vaporous refrigerant leaving the generator HX has the same temperature as the entering rich solution since it flows upwards towards the rich solution inlet (see \autoref{fig:AHPD_Process}), i.e.
\begin{equation}
	\label{eq:T_Ref_HX_G_out}
	T_{\r{Ref,GRh}}= T_{\r{RSo,SHX,out}}
 \end{equation} 
Secondly, it is assumed that the poor solution leaving the generator HX is saturated, e.g. \cite{Jeong1998,Kohlenbach2008, Evola2013, Marc2015, Sabbagh2018, Ochoa2016}. Its temperature and LiBr mass fraction therefore correlate with the high pressure via the  property  function for saturation pressure of $\r{LiBr/H_2O} $, 
\begin{equation}
	\label{eq:p_high_G}
\r{	ln}(p_{\r{high}})=\r{	ln}(p_{\r{sat,\r{LiBr/H_2O}}}(T_{\r{PSo,HX,G,out}},\xi_{\r{PSo,HX,G,out}}))
 \end{equation} 
with
\begin{equation}
	\label{eq:p_sat_LiBrH2O}
	\r{	ln}(p_{\r{sat,\r{LiBr/H_2O}}}(T,\xi))=-\r{B_{1}}+\r{B_{2}} \xi - \r{B_{3} }\xi ^2 + \r{B_{4}} T
 \end{equation}

The high pressure, however, also correlates with the condensation temperature in the condenser. Consequently, the  processes at the generator HX and the condenser HX are coupled by the high pressure. At the condenser HX surface the vaporous refrigerant is cooled down to condensation temperature, condenses and flows into the sump. 
It is common, see, e.g., \cite{Jeong1998,Kohlenbach2008, Evola2013, Marc2015, Sabbagh2018, Ochoa2016},  to assume, that the refrigerant leaves the condenser HX at saturation temperature. This means, its temperature correlates with the high pressure via the  property  function for saturation pressure of water: 
\begin{equation}
	\label{eq:p_high_C}
	\r{	ln}(p_{\r{high}})=\r{	ln}(p_{\r{sat,\r{H_2O}}}(T_{\r{Ref,HX,C,out}}))
 \end{equation} 
\begin{equation}
	\label{eq:p_sat_H2O}
	\r{	ln}(p_{\r{sat,H2O}}(T))=-\r{E_{1}}+ \r{E_{2}} T
 \end{equation} 


The transferred heat flow $ \dot{Q}_\r{C} $ at the condenser HX could now also be modeled by an approach  similar to the generator, based on $ \mathit{UA} $, but it was found that the following approach based on the HX's effectiveness $ \epsilon $ as a HX parameter resulted in a simpler AHPD model with better accuracy.  $ \epsilon $  describes the ratio between the actual and the maximum possible heat flow rate of a HX and is used, e.g., in \cite{Jeong1998, Kohlenbach2008}. 
For the condenser, the maximum possible heat flow rate would be obtained if the cooling water outlet temperature reached the temperature of the refrigerant at the HX surface, i.e., if $ T_{\r{W,AC,out}} $ equals $ T_{\r{Ref,HX,C,out}} $. Therefore  $ \dot{Q}_\r{C} $ is modeled by 
\refstepcounter{equation}\label{myeqn1}
\begin{equation}
	\label{eq:Qdot_C_epsilon}
	\dot{Q}_\r{C}=\epsilon_\r{C}~ \dot{m}_{\r{W,AC}}~ c_{p,\r{W}} \left(T_{\r{Ref,HX,C,out}}- T_{\r{W,C,in}}\right) 
 \end{equation} 

with the constant specific heat capacity $ c_{p,\r{W}} $ ($ c_{p,\r{W}} =C_2$ ) and where the effectiveness $\epsilon_\r{C}  $ is modeled as a function of the cooling water mass flow rate $ \dot{m}_{\r{W,AC}}  $:
\begin{equation}
	\label{eq:epsilon_C}
	\epsilon_\r{C}=\r{K_{C,1} }- \r{K_{C,2} }\dot{m}_{\r{W,AC}} 
 \end{equation}

Moreover, the following mass and energy balances apply:
\begin{equation}
	\label{eq:m_bal_HX_C}
	0= \dot{m}_{\r{Ref,GRh}}-\dot{m}_{\r{Ref,HX,C,out}}
 \end{equation} 
\begin{equation}
	\label{eq:E_bal_HX_C_Ref}
	0= [\dot{m}h]_{\r{Ref,GRh}} -[\dot{m}h]_{\r{Ref,HX,C,out}}- \dot{Q}_\r{C}
 \end{equation} 
\begin{equation}
		\label{eq:E_bal_HX_C_W}
	0=\dot{m}_{\r{W,AC}} (h_{\r{W,C,in}}-h_{\r{W,C,out}}) + \dot{Q}_\r{C}
 \end{equation} 

The correlation between specific enthalpies and temperatures is again described by the  property  function \autoref{eq:h_v_H2O} for vaporous and \autoref{eq:h_l_H2O} for liquid refrigerant and the cooling water.

\subsubsection{Fluid sumps}

Both in the generator and the condenser, the liquid fluid leaving the HXs flows into the corresponding fluid sump. These sumps  are approximated as continuously stirred tank reactors, i.e., their mass and energy storage capability is modeled by lumped differential mass and energy balances. 
For the generator sump, the mass and energy balances are therefore 
\begin{equation}
	\label{eq:dm_dt_G}
	\frac{\r{d}m_{\r{PSo,G}}}{\r{d}t}= \dot{m}_{\r{PSo,HX,G,out}}- \dot{m}_{\r{PSo,SHX,in}}
 \end{equation} 
\begin{equation}
		\label{eq:dm_dt_G_LiBr}
	\frac{\r{d}m_{\r{LiBr,G}}}{\r{d}t}=  [\dot{m} \xi]_{\r{PSo,HX,G,out}} - \dot{m}_{\r{PSo,SHX,in}} \frac{m_{\r{LiBr,G}}}{m_{\r{PSo,G}}} 
 \end{equation} 
\begin{equation}
			\label{eq:dE_dt_G}
	\frac{\r{d}H_{\r{PSo,G}}}{\r{d}t}=[\dot{m} h]_{\r{PSo,HX,G,out}}- \dot{m}_{\r{PSo,SHX,in}} \frac{H_{\r{PSo,G}}}{m_\r{{PSo,G}}} 
 \end{equation} 

with the stored poor solution mass and enthalpy $ m_{\r{PSo,G}} $ and  $ H_{\r{PSo,G}} $, the stored LiBr mass  $ m_{\r{LiBr,G}} $ and the mass flow rate of the poor solution flowing out of the sump $\dot{m}_{\r{PSo,SHX,in}}  $.

For the condenser sump, measurement data of the investigated AHPD showed that the mass inside the fluid sump hardly ever changes, therefore $ \dot{m}_{\r{Ref,C,out}}  $ is set equal to	$ \dot{m}_{\r{Ref,HX,C,out}}  $.
The mass and energy balances for the condenser are therefore   
\begin{equation}
	\label{eq:dm_dt_C}
	0   = \dot{m}_{\r{Ref,HX,C,out}}-\dot{m}_{\r{Ref,C,out}} 
 \end{equation} 
\begin{equation}
		\label{eq:dE_dt_C}
	\frac{\r{d}H_{\r{Ref,C}}}{\r{d}t}=  [\dot{m}h]_{\r{Ref,HX,C,out}}  - \dot{m}_{\r{Ref,C,out}} \frac{H_\r{{Ref,C}}}{m_{\r{Ref,C}}} 
 \end{equation} 
with the stored refrigerant mass and enthalpy $ m_{\r{Ref,C}} $ and  $ H_{\r{Ref,C}} $ and the mass flow rate of the refrigerant flowing out of the sump $\dot{m}_{\r{Ref,C,out}}  $.

	\subsection{Absorber and evaporator}
	\label{subsec:abs_eva}

	The modeling of the absorber and evaporator is very similar to the modeling of the generator and condenser described in the section before. Therefore, the description here will mainly be restricted to listing the equations and discussing differences where applicable. 
	
		\subsubsection{Heat exchangers}

		The heat and mass transfer at the absorber HXs is modeled by
	\begin{equation}
		\label{eq:Qdot_A}
		\dot{Q}_\r{A}=\mathit{UA}_\r{A} \overline{\Delta T}_{\r{HX,A}}
	 \end{equation} 
	\begin{equation}
		\label{eq:dT_A}
		\overline{\Delta T}_{\r{HX,A}}=\frac{T_{\r{PSo,SHX,out}}+T_{\r{RSo,HX,A,out}}}{2}-\frac{T_{\r{W,A,in}}+T_{\r{W,A,out}}}{2}
	 \end{equation} 
\begin{equation}
\mathit{UA}_\r{A} =\r{K_{A1}} + \r{K_{A2}} \dot{m}_{\r{W,AC}} + \r{K_{A3}}  \dot{m}_{\r{PSo,A,in}} + \r{K_{A4}} \dot{m}_{\r{Ref,GRl}} 
	\label{eq:UA_A}
 \end{equation} 
	\begin{equation}
		\label{eq:m_bal_HX_A}
		0= \dot{m}_{\r{PSo,A,in}}-\dot{m}_{\r{RSo,HX,A,out}} + \dot{m}_{\r{Ref,GRl}} 	
	 \end{equation} 
	\begin{equation}
				\label{eq:LiBr_bal_HX_A}
		0=  [\dot{m}\xi]_{\r{PSo,A,in}}-[\dot{m}\xi]_{\r{RSo,HX,A,out}} 
	 \end{equation} 
	\begin{equation}
				\label{eq:E_bal_HX_A_Sol}		
		0=  [\dot{m}h]_{\r{PSo,A,in}}-[\dot{m}h]_{\r{RSo,HX,A,out}} + [\dot{m}h]_{\r{Ref,GRl}} 	- \dot{Q}_\r{A}
	 \end{equation} 
	\begin{equation}
				\label{eq:E_bal_HX_A_W}				
		0=\dot{m}_{\r{W,AC}} (h_{\r{W,A,in}}-h_{\r{W,A,out}}) + \dot{Q}_\r{A}
	 \end{equation} 

One difference between the modeling of the generator and the absorber is that the solution leaving the HX of the absorber is assumed to be subcooled rather than saturated. Thus, it is assumed that the absorption process is not ideal and requires a slightly larger temperature difference. 
	To model the subcooling process, it is assumed that a small fraction of the absorber heat flow  $\phi_\r{sub}\dot{Q}_\r{A}$ accounts for subcooling the solution from saturation temperature $T_{\r{RSo,HX,A,out,sat}}$ to the actual outlet temperature  $T_{\r{RSo,HX,A,out}}$, while the rest of  $\dot{Q}_\r{A}$ accounts for the absorption process without subcooling. For the AHPD used in this paper $\phi_\r{sub}$ is estimated to be app. \SI{8}{\percent} on average by means of measurement data analysis. Note that this value merely affects steady-state accuracy and can also be set to \SI{0}{\percent} if it cannot be determined, as suggested in, e.g., \cite{Ochoa2016, Kohlenbach2008, Misenheimer2017, Evola2013, Sabbagh2018}, without noticeably affecting the model's dynamic accuracy, i.e., how well the model can reproduce the AHPD's transient behavior.
	The saturation temperature is calculated with the   property  function for saturation pressure of $\r{LiBr/H_2O}$,  \autoref{eq:p_sat_LiBrH2O}, and
	\begin{equation}
		\label{eq:p_low_A}
		\r{	ln}(p_{\r{low}})=\r{	ln}(p_{\r{sat,\r{LiBr/H_2O}}} (T_{\r{RSo,HX,A,out,sat}},\xi_{\r{RSo,HX,A,out}}))
	 \end{equation} 
	The specific enthalpy of the rich solution leaving the HX is then calculated with
	\begin{equation}
		\label{eq:h_RSo_HX_A_out}
		h_{\r{RSo,HX,A,out}}=h_{\r{RSo,HX,A,out,sat}}-\frac{  \phi_\r{sub}\dot{Q}_\r{A}}{\dot{m}_{\r{RSo,HX,A,out}}} 
	 \end{equation} 
	
	with $ h_{\r{RSo,HX,A,out,sat}}=h(T_{\r{RSo,HX,A,out,sat}}, \xi_{\r{RSo,HX,A,out}}) $ and $ h_{\r{RSo,HX,A,out}}=h( T_{\r{RSo,HX,A,out}},\xi_{\r{RSo,HX,A,out}} ) $ from \autoref{eq:h_LiBrH2O}.

		The heat and mass transfer at the evaporator HX is modeled by
\begin{equation}
	\label{eq:Qdot_E_epsilon}
		\dot{Q}_\r{E}=	\epsilon_\r{E} ~\dot{m}_{\r{W,E}} ~c_{p,\r{W}}~ \left(T_{\r{W,E,in}}-T_{\r{Ref,rec}}\right) 
	 \end{equation} 
\begin{equation}
			\label{eq:epsilon_E}
		\epsilon_\r{E}=\r{K_{E1}} -\r{K_{E2}} \dot{m}_{\r{W,E}}
	 \end{equation} 
	\begin{equation}
				\label{eq:m_bal_HX_E}
		0=\dot{m}_{\r{Ref,rec}}-\dot{m}^\r{l}_{\r{Ref,HX,E,out}}-\dot{m}^\r{v}_{\r{Ref,HX,E,out}}
	 \end{equation} 
	\begin{equation}
		\label{eq:E_bal_HX_E_Ref}
		0=[\dot{m}h]_{\r{Ref,rec}} 
		-[\dot{m}h]^\r{l}_{\r{Ref,HX,E,out}} 
		-[\dot{m}h]^\r{v}_{\r{Ref,HX,E,out}} +	\dot{Q}_\r{E}
	 \end{equation} 
	\begin{equation}
				\label{eq:E_bal_HX_E_W}
		0=\dot{m}_{\r{W,E}} (h_{\r{W,E,in}}-h_{\r{W,E,out}}) - \dot{Q}_\r{E}
	 \end{equation} 
	\begin{equation}
		\label{eq:p_low_E}
		\r{	ln}(p_{\r{low}})=\r{	ln}(p_{\r{sat,\r{H_2O}}}(T_{\r{Ref,HX,E,out}}))
	 \end{equation} 
The correlation between specific enthalpies and temperatures is again described by the  property  functions \autoref{eq:h_LiBrH2O} for $ \r{LiBr/H_2O} $, \autoref{eq:h_v_H2O} for vaporous refrigerant and \autoref{eq:h_l_H2O} for liquid refrigerant and water in the hydraulic circuits. Similarly, equation \autoref{eq:rho_H2O} for the density of water can be used to calculate mass flow rates from corresponding  volume flow rates.
	
		\subsubsection{Fluid sumps}

For the mass balance of the two remaining sumps it has to be considered that the overall mass of refrigerant and LiBr inside the AHPD remains constant. Therefore, one differential LiBr and one differential overall mass balance can be omitted and replaced by corresponding  steady-state balances over all sumps. 
For the absorber sump only the differential overall mass and energy balances are required and the transient LiBr balance is replaced by a steady-state LiBr balance: 
\begin{equation}
	\label{eq:dm_dt_A}
	\frac{\r{d}m_{\r{RSo,A}}}{\r{d}t}= [\dot{m}]_{\r{RSo,HX,A,out}} - \dot{m}_{\r{RSo,A,out}} 
 \end{equation} 
\begin{equation}
	\label{eq:LiBr_bal}
	0=m_{\r{LiBr,A}}+ m_{\r{LiBr,G}}-m_{\r{LiBr,sumps}}
 \end{equation} 
\begin{equation}
	\label{eq:dE_dt_A}
	\frac{\r{d}H_{\r{RSo,A}}}{\r{d}t}= [\dot{m}h]_{\r{RSo,HX,A,out}} - \dot{m}_{\r{RSo,A,out}} \frac{H_{\r{RSo,A}}}{m_{\r{RSo,A}}}
 \end{equation} 
where  $m_{\r{LiBr,sumps}}$ is the (constant) LiBr mass  in all sumps. 

For the evaporator sump only the transient energy balance is required and the transient mass balance can be replaced by a steady-state balance: 
\begin{equation}
	\label{eq:total_bal}
	0=m_{\r{Ref,E}}+m_{\r{Ref,C}}+m_{\r{RSo,A}}+m_{\r{PSo,G}}-m_{\r{LiBr+H_2O,sumps}}
 \end{equation} 
\begin{equation}
	\label{eq:dE_dt_E}
	\frac{\r{d}H_{\r{Ref,E}}}{\r{d}t}=  [\dot{m}h]^\r{l}_{\r{Ref,E,in}}+[\dot{m}h]^\r{l}_{\r{Ref,HX,E,out}} -\dot{m}_{\r{Ref,rec}} \frac{H_{\r{Ref,E}}}{m_{\r{Ref,E}}}
 \end{equation} 
where $m_{\r{LiBr+H_2O,sumps}}$ is the (constant)  total fluid mass (LiBr and refrigerant) in all sumps.

	\subsection{Solution heat exchanger (SHX)}
	\label{subsec:SHX}
	
	Similar to the modeling of the heat transfer in the HXs of the main components in  \autoref{subsec:gen_con} and \ref{subsec:abs_eva}, it is common to also model the heat transfer in the SHX with a constant $ \mathit{UA}$-value, e.g. \cite{Evola2013, Marc2015, Sabbagh2018}. From measurement data analysis of the investigated AHPD it was found that this approach can reproduce the steady-state behavior well, but not the dynamic  response to variations of the flow rate of rich solution, as will be shown later in \autoref{subsec:dynamic_val}. 
	Instead, it is suggested to model the heat transfer in the SHX based on the terminal temperature differences  ($ \mathit{TTD} $), also referred to as approach temperature difference, on either end of the SHX instead of modeling the transferred heat flow itself. Such a modeling approach is discussed  in \cite{Unterberger2020}, where it is used to model the operating behavior of a plate heat exchanger with varying mass flow rates. It should be mentioned that energy conservation cannot be guaranteed with this approach. However, it can reproduce the dynamic effects of volume flow variations very well and  small steady-state deviations in the energy balance can typically be tolerated in control-oriented models as long as the basic  dynamics are still represented well and the system remains stable. 
	Based on \cite{Unterberger2020} the $ \mathit{TTD} $ at steady-state for the hot end of the SHX, $ \mathit{TTD}_{\r{h}} $, can then be modeled by 
	\begin{equation}
		\label{eq:delta_T_SHX_h}
		\begin{split}
			\mathit{TTD}_{\r{h}}&=
			T_{\r{PSo, SHX,in}} -T_{\r{RSo,SHX,out,ss}} 	\\
			&=f_{\r{h}} \left([\dot{m}c_p]_{\r{RSo}},[\dot{m}c_p]_{\r{PSo}}  \right) \Delta T_{\r{SHX,in}} 
		\end{split}
	 \end{equation} 
	where $ T_{\r{RSo,SHX,out,ss}} $ corresponds to the steady-state outlet temperature of the rich solution and $  \Delta T_{\r{SHX,in}} $ to the SHX's inlet temperature difference $ (T_{\r{PSo,SHX,in}} -T_{\r{RSo,SHX,in}})  $.
	In \cite{Unterberger2020} a polynomial  of degree four is used for $f_{\r{h}} $. Here, however, a linear correlation is chosen, since the mass flow rates of rich and poor solution are usually very similar, so that a linear approximation is considered sufficient, which yields
	\begin{equation}
		f_{\r{h}} =\r{K_{h1}}+\r{K_{h2}}[\dot{m}c_p]_{\r{RSo}}-\r{K_{h3}} [\dot{m}c_p]_{\r{PSo}}
	 \end{equation} 
	Parameters  $ \r{K_{h1}}$ to $ \r{K_{h3}} $ can be determined empirically or by means of other more detailed models, such as discretized models. The specific heat capacities $ c_{p,\r{RSo}} $ and $ c_{p,\r{PSo}} $ are calculated by the corresponding  property  function:
	\begin{equation}
			\label{eq:cp_LiBrH2O}
			c_{p,\r{LiBr/H_2O}}(\xi)=\left( \frac{\partial h_{\r{LiBr/H_2O}}}{\partial T} \right)_{p,\xi}=  \r{A_4} - \r{A_5} \xi
		 \end{equation} 
	Similarly, the $ \mathit{TTD} $ at the cooler end of the SHX, $ \mathit{TTD}_{\r{c}} $, can be modeled by 
	\begin{equation}
		\label{eq:delta_T_SHX_l}
		\begin{split}
			\mathit{TTD}_{\r{c}} &=
			T_{\r{PSo,SHX,out,ss}} -T_{\r{RSo, SHX,in}} \\
			&= f_{\r{c}} \left([\dot{m}c_p]_{\r{RSo}},[\dot{m}c_p]_{\r{PSo}}  \right)	\Delta T_{\r{SHX,in}}
		\end{split}
	 \end{equation} 
	\begin{equation}
		f_{\r{c}} =	\r{K_{c1}}-\r{K_{c2}}[\dot{m}c_p]_{\r{RSo}}+\r{K_{c3}} [\dot{m}c_p]_{\r{PSo}}
	 \end{equation} 
	where $ T_{\r{PSo,SHX,out,ss}} $ corresponds to the steady-state outlet temperature of the poor solution and  $ \r{K_{c1}}$ to $ \r{K_{c3}} $ to the corresponding parameters.

\begin{figure}[h!]
		\centering
				\def\svgwidth{250pt}
				\footnotesize
		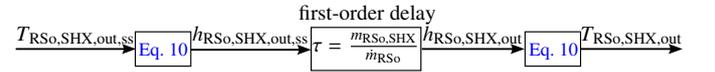
		\caption{Modeling scheme of SHX dynamics for the rich solution}
		\label{fig:SHX_dyn}%
	\end{figure}

	In order to also consider the SHX's dynamics in a simple way, the model from \cite{Unterberger2020} is extended by adding first-order  delay elements with mass-flow-dependent time constants on the rich and the poor solution side:
		\begin{equation}
		\small
		\label{eq:T_bal_SHX_RSo}
		\frac{\r{d}H_{\r{RSo,SHX}}}{\r{d}t}= \frac{\r{d}[mh]_{\r{RSo,SHX}}}{\r{d}t}=\dot{m}_{\r{RSo}} (h_{\r{RSo,SHX,out,ss}}-h_{\r{RSo,SHX}})
	 \end{equation} 
	\begin{equation}
		\small
		\label{eq:T_bal_SHX_PSo}
		\frac{\r{d}H_{\r{PSo,SHX}}}{\r{d}t}=\frac{\r{d}[mh]_{\r{PSo,SHX}}}{\r{d}t}=\dot{m}_{\r{PSo}} (h_{\r{PSo,SHX,out,ss}}-h_{\r{PSo,SHX}})
	 \end{equation} 
	
	
	where $ h_{\r{RSo,SHX,out,ss}} $ and $ h_{\r{PSo,SHX,out,ss}} $ are the specific enthalpies correlating with $ T_{\r{RSo,SHX,out,ss}} $ and $ T_{\r{PSo,SHX,out,ss}} $ from \autoref{eq:delta_T_SHX_h} and \autoref{eq:delta_T_SHX_l}, and $ h_{\r{RSo,SHX,out}} $ and $ h_{\r{PSo,SHX,out}} $ are the specific enthalpies stored in the fluid in the SHX. Again, the correlations between specific enthalpies, temperatures and mass fractions are described by the corresponding property function \autoref{eq:h_LiBrH2O}. The masses of the solution in the SHX,  $ m_{\r{RSo,SHX}} $ and  $ m_{\r{PSo,SHX}} $, can be assumed constant with sufficient accuracy. The modeling scheme of the SHX dynamics is also depicted in \autoref{fig:SHX_dyn}, for the rich solution side as an example. 

	As mentioned before, energy conservation cannot be guaranteed with this approach since it does not contain an energy balance that correlates the transferred energy on the rich and poor solution side but directly gives the two outlet temperatures instead (otherwise the model would be overdetermined).  For the investigations in this paper, the error of such a steady-state energy balance over the SHX was in the range of \SIrange{0}{12}{\%}, depending on the operating point, which is considered sufficiently small for a control-oriented model.

	\subsection{Expansion valves}
	\label{subsec:expansion_valves}
	
	In the two expansion valves (REV and SEV), the solution and refrigerant, respectively, expand from high to low pressure level.
	Since heat losses are neglected, the flow through the REV and SEV is modeled as an adiabatic process, i.e., the expansion process is assumed to be isenthalpic, which is a very common assumption in the literature on modeling of $ \r{LiBr/H_2O} $ AHPDs \cite{Jeong1998, Kohlenbach2008,  Evola2013, Marc2015,  Sabbagh2018,Ochoa2016,Zinet2012,delaCalle2016, Castro2020}. Both valves have a constant opening and are thus modeled as such.
	
	
	\subsubsection{Solution expansion valve (SEV)}
	
	
The expansion is assumed to be isenthalpic, i.e.
		\begin{equation}
			\label{eq:h_SEV}
	h_{\r{PSo,SHX,out}}=	h_{\r{PSo,A,in}}
	 \end{equation} 
	
	And assuming that the poor solution entering the SEV is sufficiently subcooled by the preceding SHX \cite{Kohlenbach2008, Jeong1998, Evola2013, Marc2015, Sabbagh2018}, no refrigerant desorbs during this process, so that the temperature before and after the SEV can be assumed to be the same, i.e.
			\begin{equation}
				\label{eq:T_SEV}
		T_{\r{PSo,SHX,out}}=	T_{\r{PSo,A,in}}
	 \end{equation} 
	
The mass flow rate through the SEV, $ \dot{m}_{\r{PSo,A,in}}  $, correlates with the pressure difference $ p_{\r{high}}-p_{\r{low}} $, the filling level in the generator and the poor solution's density. And although existing models from literature \cite{Kohlenbach2008,  Evola2013, Marc2015,  Sabbagh2018,Ochoa2016, Castro2020}  usually do consider these three influencing factors, it was found that the following much simpler approach yields satisfactory modeling accuracy too and is therefore favored over more complex approaches. 
In this paper $ \dot{m}_{\r{PSo,A,in}} $ is modeled in such a way that it linearly scales with the stored mass in the generator sump only, i.e.
\begin{equation}
	\label{eq:mdot_PSo}
	\dot{m}_{\r{PSo,A,in}} =\r{ K_{SEV} }m_{\r{PSo,G}}
 \end{equation} 
where $ \r{ K_{SEV}}  $ is determined empirically.

	
	\subsubsection{Refrigerant expansion valve (REV)}
	\label{subsubsec:REV}
	
	The mass flow rate through the REV $ \dot{m}_{\r{Ref,E,in}} $ can basically be modeled in the same way as described for the SEV in  \autoref{eq:mdot_PSo}. However, as discussed in \autoref{subsec:abs_eva},
	 measurement data of the AHPD serving as basis for the work described in this paper indicate that the level in the condenser sump hardly ever changes. Therefore $ \dot{m}_{\r{Ref,E,in}} $ and correspondingly also  $ \dot{m}_{\r{Ref,C,out}} $ is simply set equal to  $ \dot{m}_{\r{Ref,HX,C,out}}$. 
	
	As described above, also the expansion through the REV is modeled as isenthalpic process. And since the refrigerant entering the REV is liquid and saturated or at least very close to saturation, the expansion through the REV causes part of the refrigerant to evaporate, so that the stream entering the evaporator consists of a vaporous part $ \dot{m}^\r{v}_{\r{Ref,E,in}} $ and a liquid part $ \dot{m}^\r{l}_{\r{Ref,E,in}} $, which are determined by means of steady-state mass and energy balances: 
	\begin{equation}
		\label{eq:m_bal_REV}
		0= \dot{m}^\r{v}_{\r{Ref,E,in}} 
		+\dot{m}^\r{l}_{\r{Ref,E,in}} -	\dot{m}_{\r{Ref,E,in}}
	 \end{equation} 
	\begin{equation}
			\label{eq:E_bal_REV}
		0= [\dot{m}h]^\r{v}_{\r{Ref,E,in}} + [\dot{m}h]^\r{l}_{\r{Ref,E,in}} -
		\dot{m}_{\r{Ref,E,in}}\frac{H_{\r{Ref,C}}}{m_{\r{Ref,C}}}
	 \end{equation} 
	
	where $h^\r{v}_{\r{Ref,E,in}}$ and $h^\r{l}_{\r{Ref,E,in}}$ are the specific enthalpies of vaporous and liquid, saturated refrigerant.  The correlation between specific enthalpies and temperatures are again expressed by the corresponding  property  functions \autoref{eq:h_v_H2O} and \autoref{eq:h_l_H2O}. 
	and the temperature of refrigerant $T_{\r{Ref,E,in}}$ corresponds to the saturation temperature at low pressure, which is expressed by the property function for saturation pressure of water,  \autoref{eq:p_sat_H2O}.
		 
		 Note that since the predominant part of $ \dot{m}_{\r{Ref,E,in}}  $ (in general $> \SI{95}{\%}  $) flows directly into the evaporator sump, the assumption above to set  $ \dot{m}_{\r{Ref,C,out}} $ equal to  $ \dot{m}_{\r{Ref,HX,C,out}}$, i.e., to set $\frac{\r{d}m_{\r{Ref,C}}}{\r{d}t}=  0$, is always legit for AHPDs of this construction type  due to the following reason:  As already discussed in \autoref{subsec:rel_storage}, the condenser and evaporator sumps can be interpreted as two first-order delay elements in series with only the REV in between, which does not interact with any HXs or other components though. Even if the assumption $\frac{\r{d}m_{\r{Ref,C}}}{\r{d}t}= 0$ didn't apply, this would only mean that the individual time constants of the two first-order delay elements would be slightly wrong, but  the effect on their combined dynamics, and hence on the overall AHPD model, would be negligible. The same applies to AHPDs where $ \dot{m}_{\r{Ref,E,in}}  $ does not flow directly into the sump but first flows over the HX surface. In this case it would still account for a significantly smaller share compared to the recirculated $ \dot{m}_{\r{Ref,rec}} $ (in the investigated AHPD $ < \SI{5}{\%} $) and would therefore also have a negligible effect on the heat transfer at the evaporator HX.
%
	\subsection{Solution and refrigerant pump}
	\label{subsec:pumps}
	In this paper, the pump work is neglected which is a common and valid simplification \cite{Ochoa2016, Kohlenbach2008, Marc2015} for $ \r{LiBr/H2O} $ AHPD. Since the volume flow rates, and not the pump speed, are used as input variables for the models in this paper, no correlation describing the relationship between speed, pressure difference and flow rate is necessary. If required, such a correlation can be taken from e.g. \cite{Vinther2015}.

	\subsection{Overall AHPD model }
	\label{subsec:nonlin_complete}

The overall AHPD model consists of the previously discussed  \autoref{eq:Qdot_G} to \autoref{eq:E_bal_REV}. 
Note, that the property functions  \autoref{eq:rho_LiBrH2O} to \autoref{eq:h_l_H2O}, \autoref{eq:p_sat_LiBrH2O}, \autoref{eq:p_sat_H2O} and \autoref{eq:cp_LiBrH2O} occur multiple times throughout the overall model  to describe correlations between different fluid
 properties like temperature, pressure, mass fraction, specific enthalpy, and density for individual fluid streams.  
The resulting model  then consist of first-order, ordinary coupled differential equations describing the system's dynamics, and coupled algebraic equations describing, e.g., heat transfer or fluid property correlations. 
The strong coupling between the equations is reflected in the large number of model variables that occur in equations of different subsystems. From a physical point of view, the strong coupling can be explained by the fact that the process underlying AHPDs is a circular process, in which the output variables of one sub-process are in turn the input variables of one or more other sub-processes.  
Since both differential and algebraic equations are in general nonlinear, the resulting system is  a system of nonlinear differential-algebraic equations. As such it can be very useful as a simulation model to test different control strategies in simulation, but it is too complex for many model-based control approaches. 

For a simpler model structure, the model can be linearized at a reference operating point  (ROP). The ROP consists of values for all model variables and is determined using the proposed nonlinear model, by simulating the steady-state values for a set of input variables, i.e., inlet temperatures and mass flow rates in the three hydraulic circuits and the volume flow rate of rich solution. In a first step, all equations are linearized at the chosen ROP. In the next step, the system of linearized, coupled algebraic equations can  be solved and eliminated by inserting the solution into the system of differential equations, eventually yielding a model in the typical linear state-space representation: 
\begin{equation}
	\begin{split}
\dot{\v{x}}= \v{A} \v{x} + \v{B} \v{u}\\
\v{y}= \v{C} \v{x} + \v{D} \v{u}
	\end{split}
 \end{equation} 
where the state variables $ \v{x} $ are the stored masses and energies, the input variables $ \v{u} $ are the inlet temperatures and mass flow rates in the three hydraulic circuits and the volume flow rate of rich solution, the output variables $ \v{y} $ are the outlet temperatures and heat flow rates in the three hydraulic circuits, and $ \v{A} $, $ \v{B} $, $ \v{C} $ and $ \v{D} $ are the corresponding system matrices whose entries depend on the chosen ROP.  
Details on the linearization process for this AHPD model can be taken from \cite{Zlabinger2020}. 
To investigate the effect of the linearization on the model's accuracy, both the nonlinear and a linearized model version will be experimentally validated in \autoref{sec:validation}. Additionally, the new modeling approach will be compared to alternative modeling approaches as benchmark which will be explained in the next section.

\section{Benchmark modeling approaches for comparison}
\label{sec:benchmark_models}
Although some approaches from  literature on AHPD models served as the basis for the model developed in this work, it does differ in the way heat transfer is modeled in the HXs, as mentioned earlier in the corresponding sections. 
In order to compare the new modeling approaches for AHPD  HXs  to  other approaches, two benchmark model versions  are now presented in this section, in each of which either the HXs of the main components and/or the SHX is modeled in a different way compared to the AHPD model from this paper. In the following, first an overview of the model versions to be compared is given, followed by a more detailed description of the two benchmark model versions (\emph{V1} and \emph{V2} in \autoref{tab:model_version}). 



The base case
represents the AHPD model developed in this paper. It will be referred to as \emph{base-a}, and its linearized version as \emph{base-b}.
In the first benchmark model version, which will be referred to as version \emph{V1}, the HXs in the AHPD's main components generator, condenser, evaporator and absorber, are  modeled by means of constant $ \mathit{UA}$-values instead of variable $ \mathit{UA}$-values and effectiveness $ \epsilon $ as it is the case for the base model, since this is a very common approach in AHPD modeling, e.g. \cite{Jeong1998,Kohlenbach2008, Evola2013, Marc2015, Sabbagh2018}. The SHX is modeled in the same way as for the base model. 
In the second benchmark model version, which will be referred to as version \emph{V2}, also the heat transfer in the SHX is  modeled by means of a constant $ \mathit{UA} $-value  similar to AHPD models in  \cite{Evola2013, Marc2015, Sabbagh2018}  instead of  terminal temperature differences $ \mathit{TTD} $.
The modeling of the remaining AHPD components (sumps, valves, pumps) is the same for all  model versions. The model versions to be compared, including the linearized version of the base model, are summarized in  \autoref{tab:model_version}.

\begin{table}[h!]
	\centering
		\caption{Overview over different model versions to be compared}%
	\begin{tabular}{@{}ccc@{}}
		\toprule
		model version & main components' HXs               & SHX                \\ \midrule
		\emph{V1}            & constant $ \mathit{UA}$        & variable $ \mathit{TTD} $\\
		\emph{V2}               & constant $ \mathit{UA}$         & constant $ \mathit{UA}$    \\
				\emph{base-a} (nonlin.)         & variable  $ \mathit{UA}$  or $\epsilon$ & variable $ \mathit{TTD} $ \\ 
		\emph{base-b} (lin.)         & not applicable & not applicable \\ \bottomrule
	\end{tabular}
	\label{tab:model_version}
\end{table}

Specifically, the following equations are now replaced for the two benchmark model versions:
For the modeling of the generator, in versions \emph{V1} and \emph{V2}, \autoref{eq:UA_G} describing the variable $ \mathit{UA}$-value $ \mathit{UA}_\r{G} $ in dependence of the mass flow rates at the HX is replaced by a constant $\mathit{UA}_\r{G,const} $. 
 Similarly, for the modeling of the absorber, \autoref{eq:UA_A} describing the variable $\mathit{UA}_\r{A} $ is also replaced by a constant $\mathit{UA}_\r{A,const} $.
For the  modeling of the condenser \autoref{eq:Qdot_C_epsilon} and \autoref{eq:epsilon_C} describing the transferred heat flow $ \dot{Q}_\r{C} $ based on the variable  effectiveness $ 	\epsilon_\r{C} $, is replaced by
\begin{equation}
	\label{eq:Qdot_C_UA}
	\dot{Q}_\r{C}=\mathit{UA}_\r{C.const} \overline{\Delta T}_{\r{HX,C}}
 \end{equation} 
with the constant $ \mathit{UA}$-value $UA_\r{C,const}$ and  the   arithmetic mean temperature difference $\overline{\Delta T}_{\r{HX,C}}$
\begin{equation}
	\label{eq:dT_C}
	\overline{\Delta T}_{\r{HX,C}}=T_{\r{Ref,HX,C,out}}-\frac{T_{\r{W,C,in}}+T_{\r{W,C,out}}}{2}
 \end{equation} 
consisting of  the in- and outlet temperatures  of the cooling water $ T_{\r{W,C,in}} $ and $ T_{\r{W,C,out}} $ and the condensation temperature of refrigerant, i.e., the temperature of the refrigerant leaving the HX, $ T_{\r{Ref,HX,C,out}}  $. 
And finally for the modeling of the evaporator, \autoref{eq:Qdot_E_epsilon} and \autoref{eq:epsilon_E} describing the transferred heat flow $ \dot{Q}_\r{E} $ based on the variable  effectiveness $ 	\epsilon_\r{E} $, is replaced by
\begin{equation}
	\label{eq:Qdot_E_UA}
	\dot{Q}_\r{E}=\mathit{UA}_\r{E,const} \overline{\Delta T}_{\r{HX,E}}
 \end{equation} 
with the constant $ \mathit{UA}$-value $UA_\r{E,const}$ and  the  arithmetic mean temperature difference $\overline{\Delta T}_{\r{HX,E}}$
\begin{equation}
	\label{eq:dT_E}
	\overline{\Delta T}_{\r{HX,E}}= \frac{T_{\r{W,E,in}}+T_{\r{W,E,out}}}{2}-\frac{T_{\r{Ref,rec}}+T_{\r{Ref,HX,E,out}}}{2}
 \end{equation} 

In version \emph{V2}, also the modeling of the SHX is adapted. Instead of modeling the SHX by means of the terminal temperature difference approach, it is also modeled by means of a constant $ \mathit{UA}$-value $ \mathit{UA}_{\r{SHX,const}} $. Moreover, the energy storage capability of the fluid in the SHX is neglected, which are both common approaches in the literature, e.g. \cite{Evola2013, Marc2015, Sabbagh2018}.  
\autoref{eq:delta_T_SHX_h} to \autoref{eq:T_bal_SHX_PSo} are therefore replaced by 
\begin{equation}
	\label{eq:Qdot_SHX}
	\dot{Q}_\r{SHX}=UA_\r{SHX} \overline{\Delta T}_{\r{SHX}}
 \end{equation} 
with the mean HX temperature difference $\overline{\Delta T}_{\r{SHX}}$
\begin{equation}
	\label{eq:T_mean_SHX}
	\begin{split}
	\overline{\Delta T}_{\r{SHX}}=& \frac{T_{\r{PSo,SHX,in}}+T_{\r{PSo,SHX,out}}}{2}-\\
	&\frac{T_{\r{RSo,SHX,in}}+T_{\r{RSo,SHX,out}}}{2}
	\end{split}
 \end{equation} 
with the in- and outlet temperatures of the  poor solution $ T_{\r{PSo,SHX,in}}$ and $ T_{\r{PSo,SHX,out}}$ and of the rich solution $ T_{\r{RSo,SHX,in}}$ and $ T_{\r{RSo,SHX,out}}$. 

The steady-state energy balances on either side (rich and poor solution) are 
\begin{equation}
	\label{eq:E_bal_SHX_RSo}
	0=\dot{m}_{\r{RSo,SHX,in}} (h_{\r{RSo,SHX,in}}-h_{\r{RSo,SHX,out}})+\dot{Q}_\r{SHX}
 \end{equation} 
\begin{equation}
	\label{eq:E_bal_SHX_PSo}
	0=\dot{m}_{\r{PSo, SHX, in}} (h_{\r{PSo,SHX,in}}-h_{\r{PSo,SHX,out}})-\dot{Q}_\r{SHX}
 \end{equation} 
where the correlation between enthalpies, temperatures and mass fractions is again described by \autoref{eq:h_LiBrH2O}. 

These two model versions \emph{V1} and \emph{V2} will be used as benchmark for the new modeling approach in the course of the experimental validation in the next section.

	\section{Experimental validation}
	\label{sec:validation}
	
	For  validation, the new model from \autoref{sec:Sim_model} (nonlinear model \emph{base-a} in \autoref{tab:model_version}), the linearized version thereof based on the linearization procedure described in \cite{Zlabinger2020} (\emph{base-b}), and the two benchmark model versions (\emph{V1} and \emph{V2})  are implemented in MATLAB Simulink\textsuperscript{\textregistered} \cite{MathWorks2020}. Simulation results are compared to each other and to 
	measurement data from the AHPD described in \autoref{sec:process_description} to illustrate the  degree of accuracy that can be expected from these models. This will first be done for steady-state results in \autoref{subsec:static_val} and then for transient operation  in \autoref{subsec:dynamic_val}. Note that for the visualization of the validation results in this section, variables will often not be given in SI-units like \si{\kelvin} and \si{\m \cubed \per \second}, but rather in more manageable units like \si{\celsius} and \si{\m \cubed \per \hour}.
	
	The ROP for the linearized model used for this validation is listed in \autoref{tab:ROP}. This is a reasonable operating point but it is deliberately chosen to not be exactly in the middle of the investigated operating range to highlight the effect of the linearization. 
	 The constant $\mathit{UA} $-values for model version \emph{V1} and \emph{V2} are also determined based on measurement data from this operating point (see \autoref{sec:param_empirical}).

%
	\begin{table}[h!]
		\setlength{\tabcolsep}{5pt} 
		\centering
				\caption{Reference operating point (ROP) for experimental validation}%
		\begin{tabular}{ccccccc}
			\toprule
			$ T_{\r{W,G,in}}$ &  $ \dot{m}_{\r{W,G}} $ & $ T_{\r{W,AC,in}}$ & $ \dot{m}_{\r{W,AC}}$ & $ T_{\r{W,E,in}}$ & $ \dot{m}_{\r{W,E}} $ & $ \dot{V}_{\r{RSo}} $ \\
			\midrule
			\si{\celsius}	& \si{\kg \per \hour}  & \si{\celsius} &\si{\kg  \per \hour} &\si{\celsius} &\si{\kg  \per \hour}  &\si{\litre\per \hour} \\
			\midrule
			80	&1200 &  29& 6200 & 14 & 2200 & 450 \\
			\bottomrule
		\end{tabular}
		\label{tab:ROP}
	\end{table}
	
	
	\subsection{Test bench setup}
	\label{subsec:test bench}
	
	The AHPD described in \autoref{sec:process_description} is installed at a test bench that allows individual adjustment of all seven  input variables. Flow rates are adjusted by means of speed controlled pumps and measured by means of magnetic-inductive flow meters \cite{EndressHauser}, where the sensors for the flow rates in the hydraulic circuits are positioned at the AHPD outlet and the sensor for the rich solution between SHX and generator.
	 Inlet temperatures are adjusted by means of three-way-valves and in- and outlet temperatures are measured by means of Pt100 elements \cite{PMR2021}.
	As discussed in \autoref{subsec:rel_storage} the position of these temperature sensors  influences the throughput time between in- and outlet temperature sensor. Since this is due to the AHPD's hydraulic integration rather than the AHPD construction itself, this effect is not considered in the AHPD model. In order to still consider this effect for the validation, the measurement data must be adjusted accordingly. 
		For that, the measurement signals of the in- and outlet temperatures,  $ T_{\r{W,i,in,meas}} $ and $ T_{\r{W,i,out,meas}} $, are shifted by the respective dead times $ \Delta t_{\r{d,i,in}}  $ and $ \Delta t_{\r{d,i,out}} $ by
	\begin{equation}
		\label{eq:deadtime_in}
		T_{\r{W,i,in}}(t)=T_{\r{W,i,in,meas}}(t-\Delta t_{\r{d,i,in}} )
	 \end{equation} 
	\begin{equation}
				\label{eq:deadtime_out}
		T_{\r{W,i,out}}(t)=T_{\r{W,i,out,meas}}(t+\Delta t_{\r{d,i,out}} )
	 \end{equation} 
	
	where the index i denotes the subscripts G, AC and E for the hot, cooling and chilled water circuit respectively. The individual dead times are determined based on the corresponding volume flow rates and the fluid volume inside the HXs and the pipe segments between the sensors. 

	\subsection{Steady-state validation}
	\label{subsec:static_val}
	
	For the evaluation of the steady-state accuracy of the four models, all seven  input variables are varied individually and simulation results are compared to steady-state measurement data. 
	\begin{figure*}
		\centering
		\includegraphics[width=2.6 \columnwidth, angle=90]{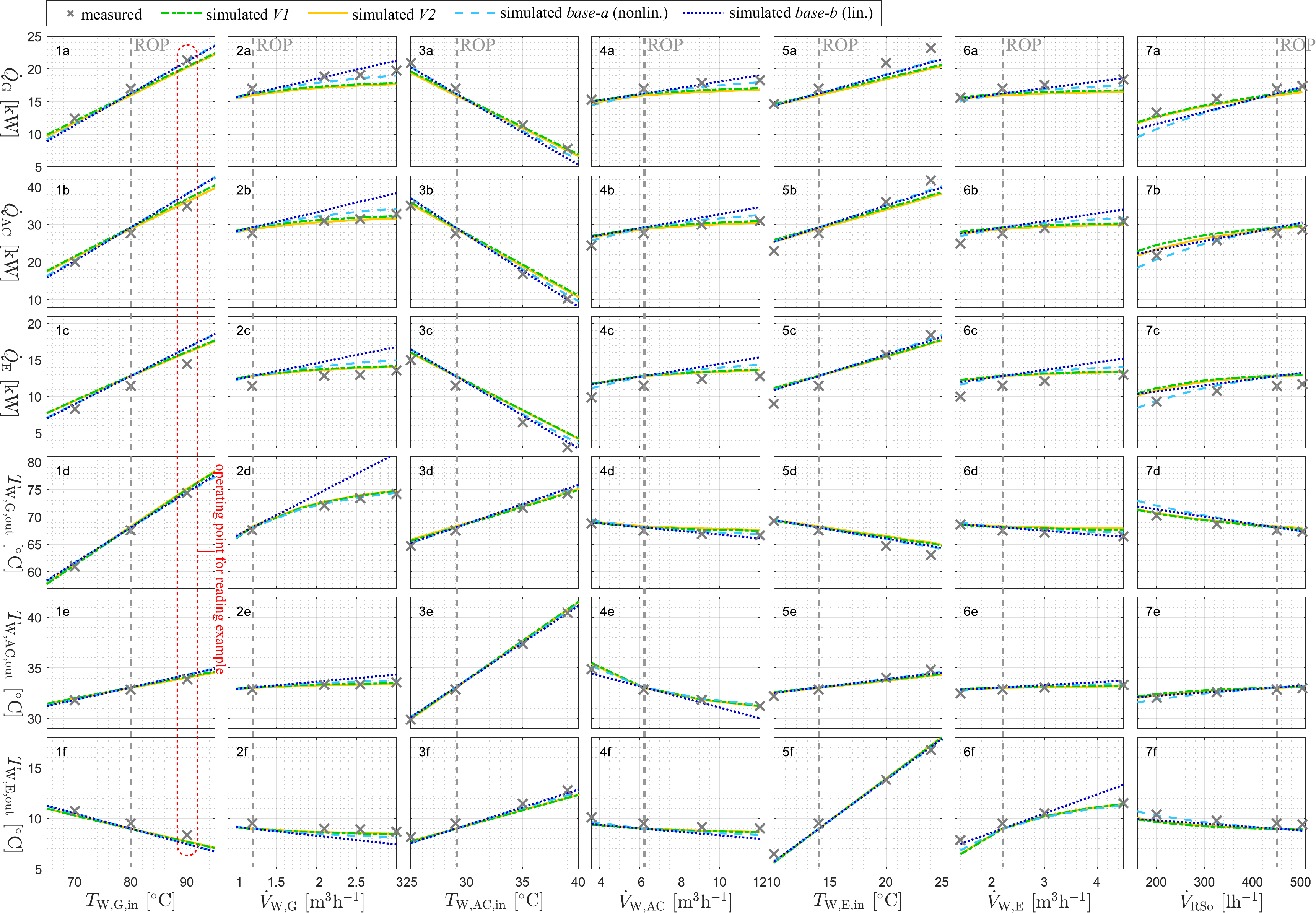}%
		\caption{Comparison of measured and simulated output variables in steady-state }%
		\label{fig:stat_val}%
	\end{figure*}
	\autoref{fig:stat_val} shows the effect of the seven  input variables $ T_{\r{W,G,in}}$,  $ \dot{V}_{\r{W,G}} $, $ T_{\r{W,AC,in}}$,  $ \dot{V}_{\r{W,AC}}$, $ T_{\r{W,E,in}}$,  $ \dot{V}_{\r{W,E}} $ and $ \dot{V}_{\r{RSo}} $, which were each varied over their entire operating range, on the three heat flow rates $ \dot{Q}_{\r{G}} $, $ \dot{Q}_{\r{AC}} $ and $ \dot{Q}_{\r{E}} $  (Subfig. 1a-7c) and the corresponding outlet temperatures $ T_{\r{W,G,out}} $, $ T_{\r{W,AC,out}} $  and  $ T_{\r{W,E,out}}$ (Subfig. 1d-7f), which are chosen as output variables. 
	The figure is arranged in such a way that each subfigure shows the correlation between one input and one output variable, when the remaining input variables are kept constant at the chosen ROP from  \autoref{tab:ROP}. Input variables are plotted on the x-axes, where the individual ROP values are marked with vertical gray dashed lines. Output variables are plotted on the y-axes. Simulation results from model version \emph{V1} are plotted in dashed-dotted green, from  \emph{V2} in solid yellow, from \emph{base-a} in dashed bright blue, from the linearized version \emph{base-b} in dotted dark blue and measurement data with gray markers. The following reading example is intended to facilitate the interpretation of the figure: For the marked opearting point (red in \autoref{fig:stat_val}) $ T_{\r{W,G,in}} = \SI{90}{\celsius}$, while the remaining six input variables have the same values as listed in \autoref{tab:ROP}. The AHPD's measured ouput variables for this operating point are 	
	\mbox{$ \dot{Q}_{\r{G}}= \SI{21.3}{ \kilo \watt} $}, 
		\mbox{$ \dot{Q}_{\r{AC}}= \SI{34.9}{ \kilo \watt} $},  
		\mbox{$ \dot{Q}_{\r{E}}= \SI{14.5}{ \kilo \watt} $}, 
		\mbox{$ T_{\r{W,G,out}} = \SI{74.4}{\celsius}$}, 
		\mbox{$ T_{\r{W,AC,out}}= \SI{33.9}{\celsius} $}  and  
		\mbox{$ T_{\r{W,E,out}}= \SI{8.4}{\celsius}$} (Subfig. 1a-1f).

	
	For variations of inlet temperatures (Subplots 1a-1f, 3a-3f and 5a-5f) it can be seen that the effects on all output variables follow a linear correlation and that the four models yield very similar simulation results, which correspond sufficiently well with measurement data.
	Correlations between volume flow rates in the hydraulic circuits and output variables  (Subfig. 2a-2f, 4a-4f, 6a-6f) are slightly nonlinear though. Accordingly, the nonlinear model versions correspond better with measurement data than the linearized model version, whose accuracy becomes worse for operating points further away from the ROP. 
	The correlations between the rich solution flow rate and output variables  (Subfig. 7a-7f) are only weakly nonlinear - the nonlinear models and the linearized one give similar results here with satisfactory accuracy. Small deviations are noticeable though. 

\begin{figure}[h!]
		\centering
		\includegraphics[width=0.85 \columnwidth]{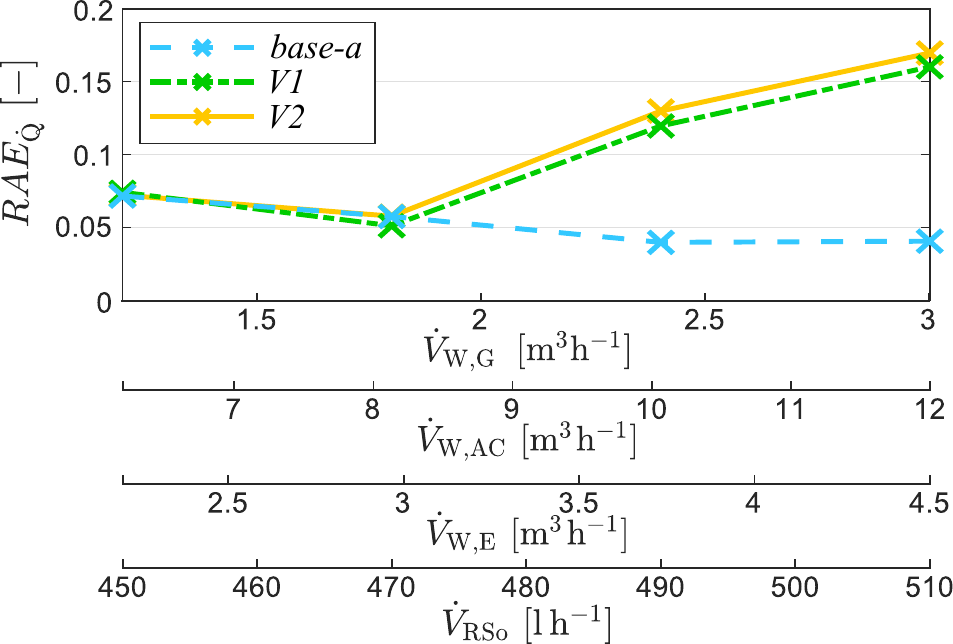}%
		\caption{Relative absolute error of heat flows ($RAE_{\dot{Q}}$) of four exemplary operating points with simultaneous variation of $\dot{V}_{\r{W,G}}$, $\dot{V}_{\r{W,AC}}$, $\dot{V}_{\r{W,E}}$ and $\dot{V}_{\r{RSo}}$ }%
		\label{fig:stat_val_2}%
	\end{figure}

	Interestingly, the model with constant $ \mathit{UA} $-values for all HXs (model version \emph{V2}) gives only slightly worse results than the model with variable heat transfer correlations (\emph{base-a}) when looking at the results in  \autoref{fig:stat_val}. If, however, operating points are considered where more than one input variable is varied from the ROP, the difference between the individual model versions becomes more apparent and \emph{base-a} yields more accurate steady-state results than \emph{V1} and \emph{V2}, which is depicted in \autoref{fig:stat_val_2}. On the y axis the relative absolute error of heat flows, $RAE_{\dot{Q}}$, is depicted which is calculated by
	\begin{equation}
		RAE_{\dot{Q}}=\frac{1}{3} \left(
		\left|\frac{\Delta\dot{Q}_{\r{G}}}{\dot{Q}_\r{G,meas}} \right|+
		\left|\frac{\Delta\dot{Q}_\r{AC}}{\dot{Q}_\r{AC,meas}} \right|+
		\left|\frac{\Delta\dot{Q}_\r{E}}{\dot{Q}_\r{E,meas}} \right| \right)
	 \end{equation} 
	where $ \Delta\dot{Q}  $ corresponds to the difference between simulated and measured heat flows and $\dot{Q}_\r{meas} $ to the measured heat flows. The diagram shows the $RAE_{\dot{Q}}$ for the three nonlinear model versions for four individual operating points, where all volume flows are varied simultaneously while the remaining input variables (inlet temperatures) are  kept constant at the ROP from \autoref{tab:ROP}. The first operating point at the left end of the scale corresponds to the volume flow rates of the ROP and the last operating point at the right end of the scale to the maximum values of the volume flow rates. It can be seen, that the $RAE_{\dot{Q}}$ of version \emph{base-a} is app. \SI{5}{\%} for all operating points but reaches app. \SI{16}{\%} and \SI{17}{\%} for \emph{V1} and \emph{V2} for the last operating point.
	The difference becomes even more apparent when extreme operating conditions are considered where very high heat flows occur. However, it should be considered that the ROP and therefore also the constant $ \mathit{UA}$-values for model versions \emph{V1} and \emph{V2} would be chosen in the middle of the expected operating range when using the model for model-based control of a real application. Therefore, also the deviations would be smaller than depicted here. Nevertheless, the base model  is able to reflect the plant behavior better in a larger operating range than model versions \emph{V1} and \emph{V2}.
	In summary, the steady-state accuracy of the nonlinear models exceeds the one of the linear model for  variations of $ \dot{V}_{\r{W,G}}$, $ \dot{V}_{\r{W,AC}}$ and $ \dot{V}_{\r{W,E}}$ and is equally good for variations of $ T_{\r{W,G,in}}$, $ T_{\r{W,AC,in}}$,  $ T_{\r{W,E,in}}$ and $ \dot{V}_{\r{RSo}} $. Among the three nonlinear model versions, the base model  yields the best results when considering the entire operating range, followed by \emph{V1} and \emph{V2}, indicating that variable heat transfer correlations yield better results than constant $ \mathit{UA} $-values. Nevertheless, all four models are considered suitable to be used for the design of advanced, model-based control strategies in terms of their steady-state accuracy.  Deviations are considered sufficiently small to be compensated by adding integral action to the model-based controller to be designed.

	\subsection{Dynamic validation}
	\label{subsec:dynamic_val}
	
	In this section measurement data from transient operating conditions are compared to corresponding simulation results from the four models. For this, three exemplary test runs are discussed, in which one input variable at a time is changed in a step-like manner, while the other  input variables are again kept constant at the selected ROP from \autoref{tab:ROP}. 
		The three test runs were chosen deliberately since they reflect the wide variety of dynamic responses to different input variable
	variations.

	\subsubsection{Step in the hot water inlet temperature $ T_{\r{W,G,in}} $}
	\label{subsubsec:T_W_G_in}
	
\begin{figure*}
		\centering
		\includegraphics[width=2.1\columnwidth, angle=0]{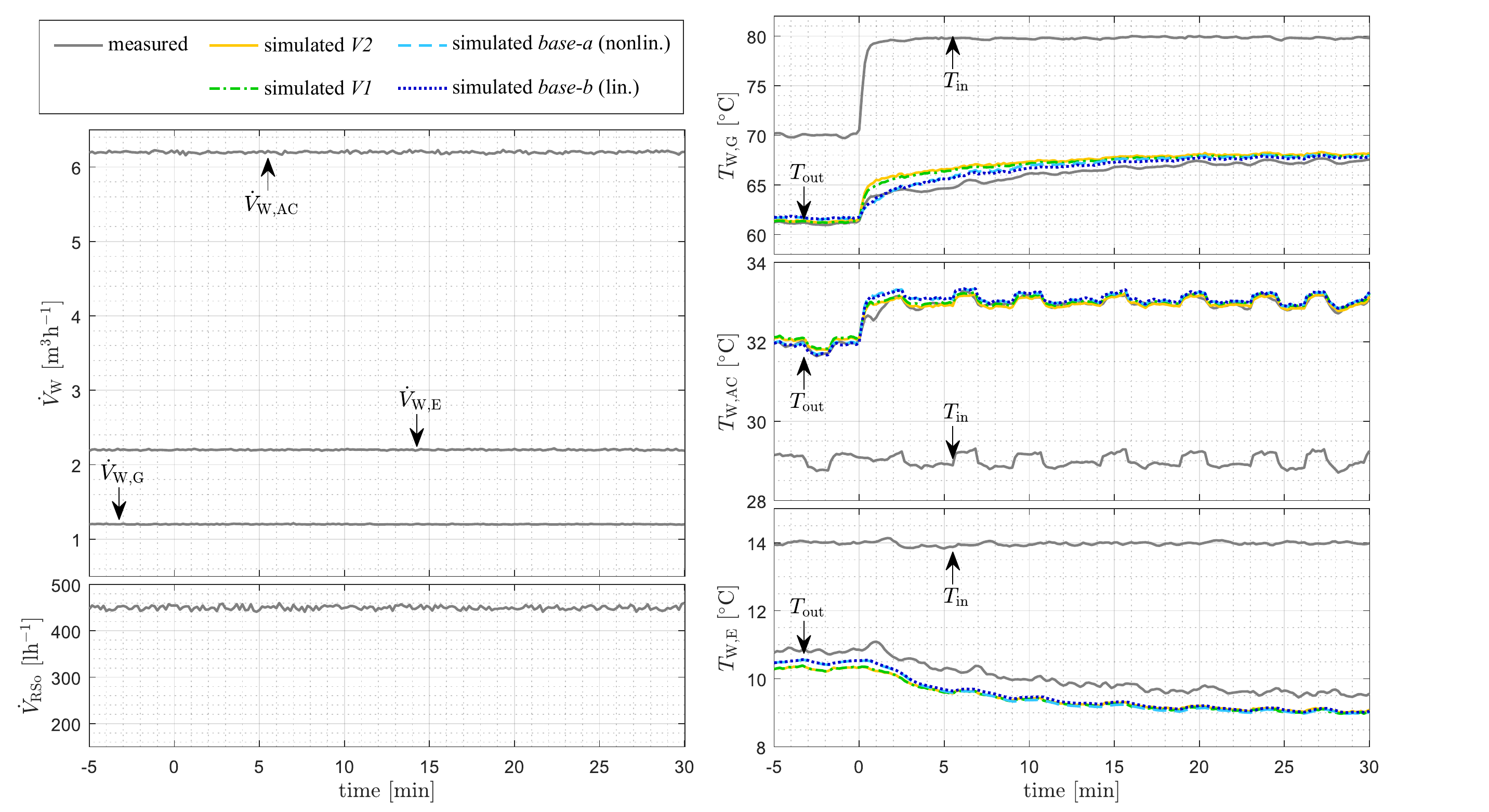}%
		\caption{Measured and simulated step response for step in hot water inlet temperature $ T_{\r{W,G,in}} $ }%
		\label{fig:step_T_W_G_in}%
	\end{figure*}

	\autoref{fig:step_T_W_G_in} shows a step-like change in the hot water inlet temperature  $ T_{\r{W,G,in}} $ and the corresponding step response for the three outlet temperatures over time. The upper left plot shows the measured volume flow rates of hot, cooling and chilled water $ \dot{V}_{\r{W,G}} $, $ \dot{V}_{\r{W,AC}} $ and $ \dot{V}_{\r{W,E}} $. The lower left plot shows the measured volume flow rate of the rich solution $ \dot{V}_{\r{RSo}} $. The plots on the right side show in- and outlet temperatures in the hot, cooling, and chilled water circuit. Outlet temperatures are again plotted in solid  gray for measured values, in dashed-dotted green  for simulated values with version \emph{V1}, in solid yellow for  \emph{V2}, in dashed bright blue for \emph{base-a} and in dotted dark blue for the linearized version \emph{base-b}. Note that the rectangular fluctuations in the cooling water inlet temperature are linked to the fact that the corresponding three-way-valve at the test bench only allows discrete valve positions.

	At $ t=\SI{0}{\minute} $ the hot water inlet temperature $ T_{\r{W,G,in}} $  is increased from $ \SI{70}{\celsius} $ to $ \SI{80}{\celsius} $. As a result the hot water  outlet temperature  $ T_{\r{W,G,out}} $ increases too. At first, the outlet temperature increases as fast as the inlet temperature but after app. \SI{30}{\s} the slope gets flatter and the new steady-state value is reached after app. \SIrange{20}{25}{\min}.
	All four models can reproduce this step response from $ T_{\r{W,G,in}} $ to $ T_{\r{W,G,out}} $  sufficiently well. 
			Due to cross-coupling effects within the AHPD also the outlet temperatures of the other two hydraulic circuits change after the step. 
	The cooling water outlet temperature $ T_{\r{W,AC,out}} $ increases sharply and quickly reaches its new steady-state value within the first minute. All four models show similar results and agree well with measurement data, although the simulated temperatures do increase slightly faster than the measured one. 
		The chilled water outlet temperature $ T_{\r{W,E,out}} $ starts decreasing app. \SI{45}{\s}  after the step and reaches steady-state after app. \SI{30}{\min}, which is also reproduced well by all four models.	The steady-state difference between simulated and measured values of $ T_{\r{W,E,out}} $ is considered acceptable for the use in model-based control, as already discussed in  \autoref{subsec:static_val}. 

	\subsubsection{Step in the chilled water volume flow rate $ \dot{V}_{\r{W,E}} $}
	\label{subsubsec:Vdot_W_E}
\begin{figure*}
		\centering
		\includegraphics[width=2.1\columnwidth, angle=0]{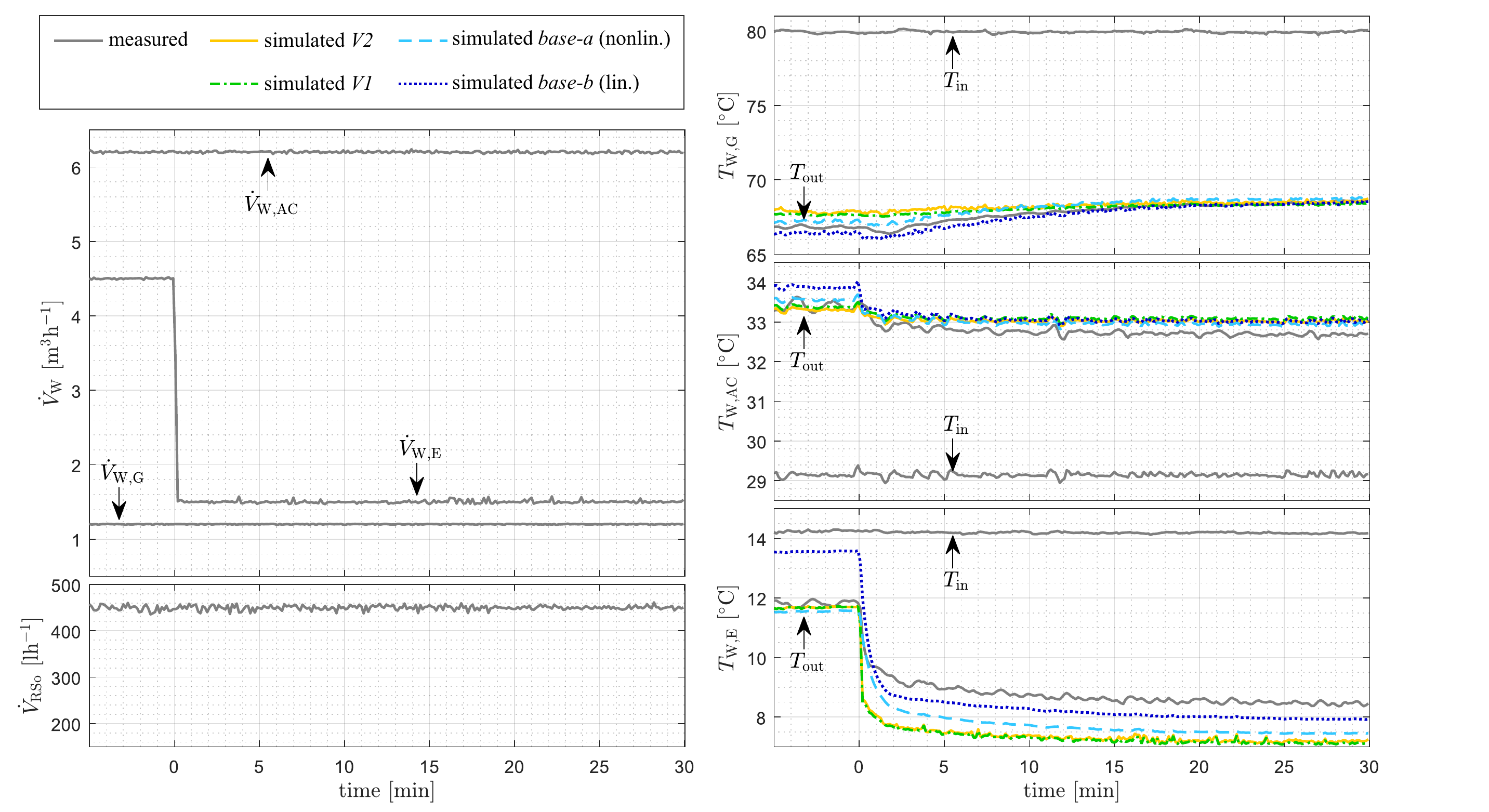}%
		\caption{Measured and simulated step response for step in chilled water volume flow rate $ \dot{V}_{\r{W,E}} $ }%
		\label{fig:Vdot_W_E}%
	\end{figure*}

	\autoref{fig:Vdot_W_E} shows a test run similar to the one described before. Here, the chilled water volume flow rate  $ \dot{V}_{\r{W,E}} $ is changed in a step-like manner from $ \SI{4,5}{\meter \cubed \per \hour} $ to $   \SI{1,5}{\meter \cubed \per \hour}$. 
	After the step $ T_{\r{W,E,out}} $ first decreases sharply and then slowly settles at its new steady-state value after app. \SI{25}{\min}. 
	 $ T_{\r{W,AC,out}} $ also decreases sharply within the first minute and changes only slightly afterwards. 
	In the hot water circuit, $ T_{\r{W,G,out}} $ first slightly decreases and then increases to  reach its new steady-state value after app.  \SI{30}{\min}. All models describe this transient behavior well. It can also be seen that steady-state deviations are higher for operating points that are further away from the ROP for the linearized model \emph{base-b}, as discussed in  \autoref{subsec:static_val}.
	
	 Furthermore, especially in the first part of the test run, before the step, it can be seen that the measured cooling and chilled water outlet temperatures $ T_{\r{W,AC,out}} $ and $ T_{\r{W,E,out}} $ slightly oscillate with a period time of app. \SI{2}{\min}. This is a phenomenon that has been observed in some test runs with this AHPD and is not caused by fluctuating input variables, but by a fluctuating poor solution flow rate, which in return is probably related to rising gas bubbles inside the AHPD. From a large number of test runs, however, it was found that these oscillations never lead to instability or other serious operating problems.
	In the test run shown in  \autoref{fig:Vdot_W_E}, the oscillations were more intense at the first operating point (before the step) and subsided after the step. Since this phenomenon is very complex and its effect on the outlet temperatures are considered to be sufficiently small, it is not considered in any  of the models.

	\subsubsection{Step in the rich solution volume flow rate $ \dot{V}_{\r{RSo}} $ }
	
\begin{figure*}
		\centering
		\includegraphics[width=2.1\columnwidth, angle=0]{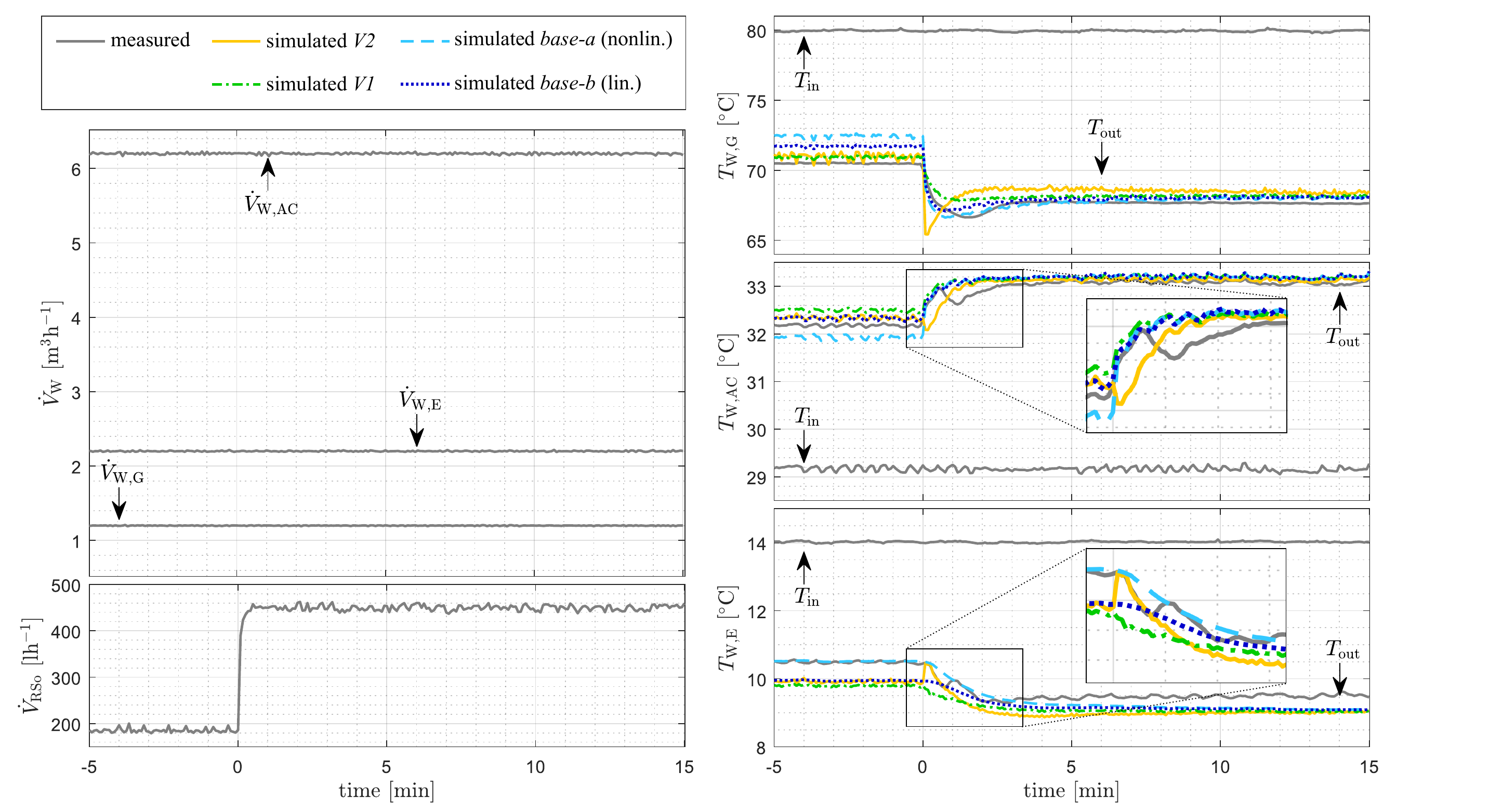}%
		\caption{Measured and simulated step response for step in rich solution volume flow rate $ \dot{V}_{\r{RSo}} $}%
		\label{fig:Vdot_RSo_step}%
	\end{figure*}

	
	\autoref{fig:Vdot_RSo_step} shows a test run where the rich solution volume flow rate $ \dot{V}_{\r{RSo}} $ is changed in a step-like manner from $ \SI{185}{\litre \per \hour} $ to $   \SI{450}{\litre \per \hour} $. 
	After the step the measured $ T_{\r{W,G,out}} $ first decreases relatively fast and then rises again slightly to reach the new steady-state value after app. \SI{3}{\min}. 
	Unlike the previously discussed test runs, the models now provide relatively different results for this step: While the model versions \emph{base-a}, \emph{base-b} and \emph{V1}  (with variable $ \mathit{TTD} $ and dynamics in the SHX model) are able to reproduce the main dynamics well, the shape of the step response from \emph{V2} (with constant $ \mathit{UA} $ and no dynamics in the SHX model)  does not fit the measured curve well, since \emph{V2}  gives a temperature change in the wrong direction for $ T_{\r{W,AC,out}} $ and $ T_{\r{W,E,out}} $, which can lead to serious control performance degradation (e.g. oscillations or instability) if such a model is used for dynamic model based control strategies. 
	
\begin{figure}[h!]
		\centering
		\includegraphics[width=0.95\columnwidth, angle=0]{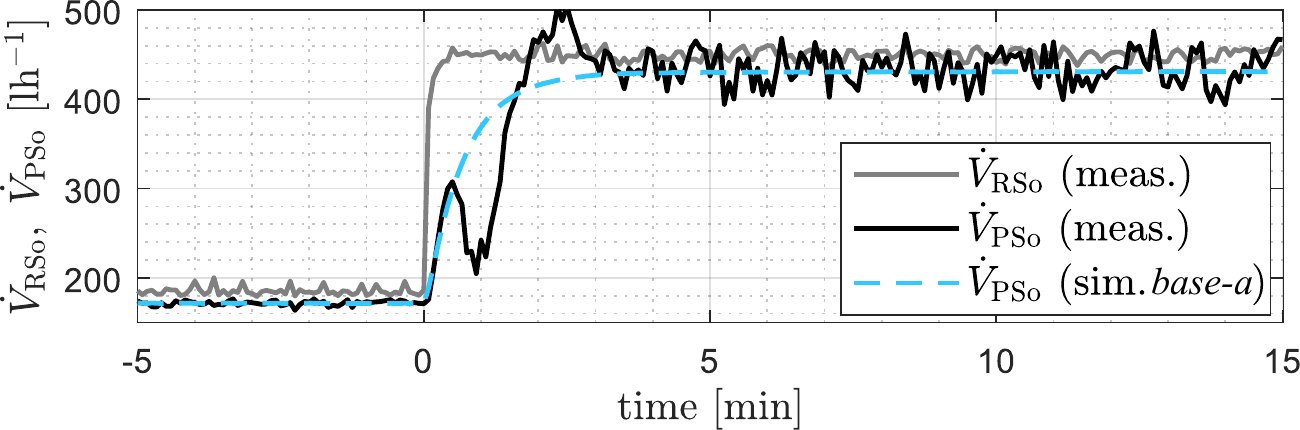}%
		\caption{Rich and poor solution volume flow rate after step in  $ \dot{V}_{\r{RSo}} $}%
		\label{fig:Vdot_PSo}%
	\end{figure}

However, also  \emph{base-a}, \emph{base-b} and \emph{V1}  show a flaw when comparing measured and simulated $ T_{\r{W,AC,out}} $:  After the step, the measured $ T_{\r{W,AC,out}} $ first increases fast, then a pronounced temperature drop can be seen after app. \SI{30}{\s} before it increases again and reaches the new steady-state value after app. \SI{5}{\min}. 
The models do reproduce the main dynamics well, but not the temperature drop shortly after the step. Taking a closer look at the solution circuit, it becomes clear that this temperature drop is linked to a drop in the poor solution flow rate:  \autoref{fig:Vdot_PSo} shows the measured rich and poor solution flow rate (measured at the entry of the generator and absorber respectively), as well as the simulated poor solution volume flow rate (using  \autoref{eq:rho_LiBrH2O} to calculate $ \dot{V}_{\r{PSo}} $ from the simulated $ \dot{m}_{\r{PSo}} $) during the discussed test run. For the sake of better visualization, only the simulation results of \emph{base-a} are shown since the other models yield virtually identical results for $ \dot{V}_{\r{PSo}} $. 
	It can be seen that the measured $ \dot{V}_{\r{PSo}} $ significantly drops at the same time as $ T_{\r{W,AC,out}} $ drops. During this drop, less poor solution arrives in the absorber, which leads to a decrease of  absorber performance, which in turn is reflected in the drop in  $ T_{\r{W,AC,out}} $. 
	The drop in $ \dot{V}_{\r{PSo}} $ is again assumed to be connected to gas bubbles rising from the absorber to the generator under these operating conditions which is not considered in any of the models. Therefore, 
	the corresponding  drop in $ T_{\r{W,AC,out}} $ cannot be reproduced either. 
	The main dynamics (the time constant of the step responses in \autoref{fig:Vdot_RSo_step}) are still represented well by \emph{base-a}, \emph{base-b} and \emph{V1}  though and such sudden, large changes in the rich solution flow rate are unlikely to occur under realistic operation conditions. Nevertheless, during control design  it can still be useful to consciously consider this deviation 
	between the modeled and the real system 
	and limit the rate of change for the rich solution flow rate if it is used as a manipulated variable. 
	
	
	
	
	It should be mentioned that the authors also investigated a modeling approach with constant $ \mathit{UA} $ in the SHX as in version \emph{V2} in combination with additional first-order delay elements as in version \emph{base-a} and \emph{V1}. Although the dynamic modeling accuracy does improve compared to \emph{V2}, the basic problem of first yielding temperature changes in the wrong direction for $ T_{\r{W,AC,out}} $ and $ T_{\r{W,E,out}} $ remains. 
	Therefore, for the modeling of the SHX the approach based on variable $ \mathit{TTD} $ should be favored over the approach with constant $ \mathit{UA} $ if $ \dot{V}_{\r{RSo}} $ is expected to vary. 
	
	
	\subsubsection{Steps in other input variables}
	The authors also investigated step responses for variations in the remaining four  input variables and observed similar dynamic accuracy as discussed above. 
	Furthermore, it was found that time constants for variations in the cooling water circuit are much smaller ($<$ \SI{1}{\min}) than for the hot and chilled water circuit.  This is explained by the fact that the cooling water circuit is connected to two components - to the absorber and to the condenser - and therefore,
	their effects show much faster throughout the whole AHPD.

	%
	

	%

	\section{Conclusion and outlook}
	\label{sec:conclusion}
	The aim of this paper was to find a model suitable to be used for the design of dynamic model-based control of AHPDs. First, an assessment of mass and energy stores showed that the most relevant dynamic effects are caused by the fluid in the sumps and in the SHX. Three model versions, differing in the way the heat transfer is modeled in the main components' HXs on the one hand and in the SHX on the other hand,  were then presented and investigated. The experimental validation results indicated that models with variable heat transfer correlations (model version \emph{base-a}) yield better steady-state accuracy over a wider operating range than models that use constant $ \mathit{UA} $-values in the main components' HXs (version \emph{V1} and \emph{V2} ). 
	Furthermore, it was shown that the common approach to model heat transfer in the SHX with constant $ \mathit{UA} $-values (\emph{V2}) does provide satisfactory steady-state accuracy and can also represent dynamic load changes of almost all input variables well but it cannot model dynamic changes of the rich solution volume flow rate $ \dot{V}_\r{RSo} $  appropriately. Instead,  it is recommended to model heat transfer in the SHX by means of an approach based on terminal temperature differences (as used in \emph{base-a} and \emph{V1}).
	
	Experimental validation also revealed a flaw that affects all model versions: For the step-like change in  $ \dot{V}_\r{RSo} $ a fluctuation in the poor solution flow rate $ \dot{V}_\r{PSo} $ - probably caused by rising gas bubbles - was observed in the investigated AHPD. None of the models consider this phenomenon, which should be taken into account when using $ \dot{V}_\r{RSo} $  as manipulated variable in a later control problem. 
	
	Another interesting finding of this contribution is that the linearization of the presented model
	only slightly impairs the model's steady-state accuracy for variations of the volume flow rates in the hydraulic circuits, but not significantly for variations of the other input variables. Additionally, the dynamic accuracy, i.e., how well the system's main dynamics can be reproduced, appears to be almost unaffected by the linearization process so that the linearized model  can be considered highly suitable to be used for model-based control. It has the very simple mathematical structure of a linear state-space model and can describe dynamic and steady-state effects of variations of all input variables sufficiently well.

	As a next step, the scalability of the suggested models shall be investigated for AHPDs of different sizes.
	First investigations showed that it should be possible to re-parameterize the model with only basic information like cooling capacity in at least one operating point as well as refrigerant and LiBr filling amount. 
	Moreover, the authors plan to use the linearized model  to design a model-based controller to extend the operating range of AHPDs, and in the course of this, to use the developed nonlinear model  as a simulation model for a virtual test bench to test different control settings. 
	
	

	\section{Acknowledgments}
%
	The research leading to these results received funding from the Austrian Climate and Energy Fund under the Grant No. 865095  in the framework of the \textit{Energieforschungsprogramm 2017} program, from the Horizon 2020 program under Grant No. 792276 and from the COMET program, which is managed by the Austrian Research Promotion Agency (FFG) and is co-financed by the Republic of
	Austria and the Federal Provinces of Vienna, Lower Austria and Styria.
	
	
	\bibliographystyle{elsarticle-num} 
	\bibliography{Lit_ModellingPaper}
	
	\newpage
	\appendix

	\section{Empirical parameters for submodels}
	\label{sec:param_empirical}
	
	\autoref{tab:param_empirical} summarizes the empirically determined parameters for the modeling approaches in \autoref{sec:Sim_model} and \autoref{sec:benchmark_models}. 
	
	\begin{table}[H]
		\centering
				\caption{Empirical parameters for submodels}%
		\begin{tabular}{lll}
			\toprule
			submodel                                    & \multicolumn{2}{c}{parameter}       \\
			\midrule
			$\mathit{UA}_{\r{G}}$, const. for \emph{V1} and \emph{V2} & $\mathit{UA}_{\r{G,const.}}$   & 2.594E+03 \\
			\midrule
			$\mathit{UA}_{\r{C}}$, const. for \emph{V1} and \emph{V2} & $\mathit{UA}_{\r{C,const.}}$   & 1.731E+03 \\
			\midrule
			$\mathit{UA}_{\r{E}}$, const. for \emph{V1} and \emph{V2} & $\mathit{UA}_{\r{E,const.}}$   & 4.337E+03 \\
			\midrule
			$\mathit{UA}_{\r{A}}$, const. for \emph{V1} and \emph{V2} & $\mathit{UA}_{\r{A,const.}}$   & 7.759E+03 \\
			\midrule
			$\mathit{UA}_{\r{SHX}}$, const. for \emph{V2}      & $\mathit{UA}_{\r{SHX}}$ & 2.803E+03 \\
			\midrule
			$\mathit{UA}_{\r{G}}$, for base model               & $\r{K_{G1}}$            & 2.523E+02 \\
			& $\r{K_{G2}}$            & 7.238E+02 \\
			& $\r{K_{G3}}$            & 5.003E+03 \\
			& $\r{K_{G4}}$            & 1.642E+05 \\
			\midrule
			$\epsilon_{\r{C}}$, for base model                  & $\r{K_{C1}}$            & 8.836E-01 \\
			& $\r{K_{C2}}$            & 1.034E-01 \\
			\midrule
			$\epsilon_{\r{E}}$, for base model            & $\r{K_{E1}}$            & 1.046E+00 \\
			& $\r{K_{E2}}$            & 1.964E-01 \\
			\midrule
			$\mathit{UA}_{\r{A}}$, for base model           & $\r{K_{A1}}$            & 1.673E+03 \\
			& $\r{K_{A2}}$            & 2.483E+02 \\
			& $\r{K_{A3}}$            & 1.812E+03 \\
			& $\r{K_{A4}}$            & 6.038E+04 \\
			\midrule
			$\mathit{TTD}_{\r{h}}$, for \emph{V1} and base model       & $\r{K_{h1}}$          & 1.109E-02 \\
			& $\r{K_{h2}}$          & 7.704E-04 \\
			& $\r{K_{h3}}$          & 4.409E-04 \\
			\midrule
			$\mathit{TTD}_{\r{l}}$, for \emph{V1} and base model    & $\r{K_{l1}}$          & 7.319E-02 \\
			& $\r{K_{l2}}$          & 2.637E-04 \\
			& $\r{K_{l3}}$          & 4.150E-04 \\
			\midrule
			$\dot{m}_{\r{PSo,A,in}}$                    & $K_\r{SEV}$               & 2.579E-02   \\
			\bottomrule
		\end{tabular}
		\label{tab:param_empirical}
	\end{table}

	\section{Parameters for mass stores}
	\label{sec:param_mass}
	
	\autoref{tab:mass} summarizes stored fluid masses  used  for the modeling approaches in \autoref{sec:Sim_model} and \autoref{sec:benchmark_models}. 
	
	\begin{table}[H]
		\centering
				\caption{Stored fluid masses}%
		\begin{tabular}{lc}
			\bottomrule
			fluid                                       & mass [kg] \\
			\midrule
		  LiBr in sumps $m_{\r{LiBr,sumps}}$   & 15.98     \\
			$\r{H_2O}$ and LiBr in sumps $m_{\r{LiBr+H_2O,sumps}}$  & 49.36   \\
		refrigerant in condenser sump $m_{\r{Ref,C}}$    & 4.15    \\
			rich solution in SHX $m_{\r{RSo,SHX}}$     & 5.32        \\
		poor solution in SHX $m_{\r{PSo,SHX}}$  & 5.32     \\
			\bottomrule
		\end{tabular}
		\label{tab:mass}
	\end{table}

	
	\newpage
	
	\section{Parameters for  property  functions}
	\label{sec:param_substance_prop}
	
	\autoref{tab:substance} summarizes the parameters for the used  property  functions and the range for mass fraction of $ \r{LiBr} $  in the solution $ \xi ~[-] $, temperature $ T~[\si{\kelvin}] $ and pressure $ p~[\si{\pascal}]$, that were used for parametrization. 
	The parameters for functions for saturation pressure  ($\r{ln}(p_{\r{sat,\r{LiBr/H_2O}}})$ and $\r{ln}(p_{\r{sat,\r{H_2O}}})$) are given for two individual ranges, because different parameters are used for high and low pressure components. 
	As discussed in \autoref{sec:Sim_model} all  property  functions used in this paper are simplified functions based on more complex  property  functions from  \cite{Yuan2005}  for $ \r{LiBr/H_2O} $ and from \cite{Wagner2002} for water. 

	\begin{table}[H]
				\caption{Parameters for  property  functions}%
		\begin{tabular}{lcll}
			\toprule
			property                            & range                           & \multicolumn{2}{c}{parameters} \\
			\midrule
			$ h_{\r{LiBr/H_2O}} $               & $ 0.45<\xi< 0.6$                & $\r{A_1}$     & 6.892E+05     \\
			& $ 293.15<T< 363.15$             & $\r{A_2}$      & 7.001E+05     \\
			&                                 & $\r{A_3}$      & 1.738E+06     \\
			&                                 & $\r{A_4}$      & 3.617E+03     \\
			&                                 & $\r{A_5}$      & 2.827E+03     \\
			\midrule
			$\r{ln}(p_{\r{sat,\r{LiBr/H_2O}}})$ & $ 0.45<\xi< 0.6$                & $\r{B_1}$      & 1.226E+01     \\
			& $ 293.15<T< 328.15$             & $\r{B_2}$      & 1.042E+01     \\
			& $p< 2200$                       & $\r{B_3}$      & 1.944E+01     \\
			&                                 & $\r{B_4}$      & 6.237E-02     \\
			\cmidrule{2-4}
			& $ 0.45<\xi< 0.6$                & $\r{B_1}$      & 6.804E+00     \\
			& $ 323.15<T< 363.15$             & $\r{B_2}$      & 7.405E+00     \\
			& $5000<p< 14000$                 & $\r{B_3}$      & 1.483E+01     \\
			&                                 & $\r{B_4}$      & 4.647E-02     \\
			\midrule
			$ \rho_{\r{LiBr/H_2O}} $            & $ 0.45<\xi< 0.6$                & $\r{R_1}$      & 1.349E+03     \\
			& $ 293.15<T< 363.15$             & $\r{R_2}$      & 2.274E+02     \\
			&                                 & $\r{R_3}$      & 1.856E+03     \\
			&                                 & $\r{R_4}$      & 5.569E-01     \\
			\midrule
			$ h^{\r{l}}_{\r{H_2O}} $            & $ 278.15<T< 368.15$             & $\r{C_1}$      & 1.143E+06     \\
			&                                 & $\r{C_2}$      & 4.186E+03     \\
			\midrule
			$ h^{\r{v}}_{\r{H_2O}} $            & $ T_{\r{sat}}<T< T_{\r{sat}}+5$ & $\r{D_1}$      & 2.009E+06     \\
			& $800<p< 15000$                  & $\r{D_2}$      & 1.803E+03     \\
			\midrule
			$\r{ln}(p_{\r{sat,\r{H_2O}}})$      & $ 277.15<T< 293.15$             & $\r{E_1}$      & 1.158E+01     \\
			&                                 & $\r{E_2}$      & 6.599E-01     \\
			\cmidrule{2-4}
			& $ 303.15<T< 323.15$             & $\r{E_1}$      & 7.591E+00     \\
			&                                 & $\r{E_2}$      & 5.266E-02     \\
			\midrule
			$ \rho^{\r{l}}_{\r{H_2O}} $         & $ 278.15<T< 368.15$             & $\r{F_1}$      & 7.397E-01     \\
			&                                 & $\r{F_2}$      & 1.984E-03     \\
			&                                 & $\r{F_3}$      & 3.760E-06     \\       
			\bottomrule
		\end{tabular}
		\label{tab:substance}
	\end{table}

\end{document}

%% file: pics/AHPS_modelling_V6b.pdf_tex
\begingroup%
  \makeatletter%
  \providecommand\color[2][]{%
    \errmessage{(Inkscape) Color is used for the text in Inkscape, but the package 'color.sty' is not loaded}%
    \renewcommand\color[2][]{}%
  }%
  \providecommand\transparent[1]{%
    \errmessage{(Inkscape) Transparency is used (non-zero) for the text in Inkscape, but the package 'transparent.sty' is not loaded}%
    \renewcommand\transparent[1]{}%
  }%
  \providecommand\rotatebox[2]{#2}%
  \newcommand*\fsize{\dimexpr\f@size pt\relax}%
  \newcommand*\lineheight[1]{\fontsize{\fsize}{#1\fsize}\selectfont}%
  \ifx\svgwidth\undefined%
    \setlength{\unitlength}{509.1221895bp}%
    \ifx\svgscale\undefined%
      \relax%
    \else%
      \setlength{\unitlength}{\unitlength * \real{\svgscale}}%
    \fi%
  \else%
    \setlength{\unitlength}{\svgwidth}%
  \fi%
  \global\let\svgwidth\undefined%
  \global\let\svgscale\undefined%
  \makeatother%
  \begin{picture}(1,0.59181993)%
    \lineheight{1}%
    \setlength\tabcolsep{0pt}%
    \put(0,0){\includegraphics[width=\unitlength,page=1]{AHPS_modelling_V6b.pdf}}%
    \put(0.65191693,0.05758395){\color[rgb]{1,1,1}\makebox(0,0)[t]{\lineheight{1.25}\smash{\begin{tabular}[t]{c}$m_{\r{RSo,A}}$\end{tabular}}}}%
    \put(0,0){\includegraphics[width=\unitlength,page=2]{AHPS_modelling_V6b.pdf}}%
    \put(0.24900032,0.55911267){\color[rgb]{0,0,0}\makebox(0,0)[lt]{\lineheight{1.70454276}\smash{\begin{tabular}[t]{l}condenser (C)\end{tabular}}}}%
    \put(0.40382862,0.47440122){\color[rgb]{0,0,0}\makebox(0,0)[lt]{\lineheight{1.70454276}\smash{\begin{tabular}[t]{l}heat exchanger (HX)\end{tabular}}}}%
    \put(0.40396583,0.43833518){\color[rgb]{0,0,0}\makebox(0,0)[lt]{\lineheight{1.70454276}\smash{\begin{tabular}[t]{l}sump\end{tabular}}}}%
    \put(0.9186812,0.50709237){\color[rgb]{0,0,0}\makebox(0,0)[t]{\lineheight{1.70454276}\smash{\begin{tabular}[t]{c}hot water \end{tabular}}}}%
    \put(0.91623899,0.48273882){\color[rgb]{0,0,0}\makebox(0,0)[t]{\lineheight{1.70454276}\smash{\begin{tabular}[t]{c}circuit\end{tabular}}}}%
    \put(0.78708168,0.55913378){\color[rgb]{0,0,0}\makebox(0,0)[rt]{\lineheight{1.70454276}\smash{\begin{tabular}[t]{r}generator (G)\end{tabular}}}}%
    \put(0.79257073,0.52277519){\color[rgb]{0,0,0}\makebox(0,0)[lt]{\lineheight{1.70454073}\smash{\begin{tabular}[t]{l}$\dot{m}_{\r{W,G,out}}$\end{tabular}}}}%
    \put(0.68164607,0.45345523){\color[rgb]{0,0,0}\makebox(0,0)[lt]{\lineheight{1.70454073}\smash{\begin{tabular}[t]{l}$\dot{m}_{\r{PSo,HX,G,out}}$\end{tabular}}}}%
    \put(0.79453619,0.47256115){\color[rgb]{0,0,0}\makebox(0,0)[lt]{\lineheight{1.70454073}\smash{\begin{tabular}[t]{l}$\dot{m}_{\r{W,G,in}}$\end{tabular}}}}%
    \put(0.24660845,0.52266003){\makebox(0,0)[rt]{\lineheight{1.25}\smash{\begin{tabular}[t]{r}$\dot{m}_{\r{W,C,out}}$\end{tabular}}}}%
    \put(0.24660846,0.47351084){\makebox(0,0)[rt]{\lineheight{1.25}\smash{\begin{tabular}[t]{r}$\dot{m}_{\r{W,C,in}}$\end{tabular}}}}%
    \put(0.51987569,0.51386753){\makebox(0,0)[t]{\lineheight{1.25}\smash{\begin{tabular}[t]{c}$\dot{m}_{\r{Ref,GRh}}$\end{tabular}}}}%
    \put(0.34936405,0.43755141){\makebox(0,0)[rt]{\lineheight{1.25}\smash{\begin{tabular}[t]{r}$\dot{m}_{\r{Ref,HX,C,out}}$\end{tabular}}}}%
    \put(0.6538101,0.40294233){\makebox(0,0)[t]{\lineheight{1.25}\smash{\begin{tabular}[t]{c}$m_{\r{PSo,G}}$\end{tabular}}}}%
    \put(0.7296254,0.49711249){\color[rgb]{0,0,0}\makebox(0,0)[lt]{\lineheight{1.25}\smash{\begin{tabular}[t]{l}$\dot{Q}_{\r{G}}$\end{tabular}}}}%
    \put(0.30468351,0.49912026){\makebox(0,0)[rt]{\lineheight{1.25}\smash{\begin{tabular}[t]{r}$\dot{Q}_{\r{C}}$\end{tabular}}}}%
    \put(0,0){\includegraphics[width=\unitlength,page=3]{AHPS_modelling_V6b.pdf}}%
    \put(0.43609622,0.55915146){\color[rgb]{0,0,0}\makebox(0,0)[lt]{\lineheight{1.70454276}\smash{\begin{tabular}[t]{l}high pressure gas room (GRh)\end{tabular}}}}%
    \put(0,0){\includegraphics[width=\unitlength,page=4]{AHPS_modelling_V6b.pdf}}%
    \put(0.41917382,0.32853429){\color[rgb]{0,0,0}\makebox(0,0)[rt]{\lineheight{1.70454276}\smash{\begin{tabular}[t]{r}refrigerant expansion\\\end{tabular}}}}%
    \put(0.41996502,0.30599617){\color[rgb]{0,0,0}\makebox(0,0)[rt]{\lineheight{1.70454276}\smash{\begin{tabular}[t]{r}valve (REV)\\\end{tabular}}}}%
    \put(0.70246837,0.25913506){\color[rgb]{0,0,0}\makebox(0,0)[lt]{\lineheight{1.70454276}\smash{\begin{tabular}[t]{l}solution expansion valve (SEV)\\\end{tabular}}}}%
    \put(0.70246837,0.33343646){\color[rgb]{0,0,0}\makebox(0,0)[lt]{\lineheight{1.70454276}\smash{\begin{tabular}[t]{l}solution heat exchanger (SHX)\\\end{tabular}}}}%
    \put(0.5779389,0.25913506){\color[rgb]{0,0,0}\makebox(0,0)[rt]{\lineheight{1.70454276}\smash{\begin{tabular}[t]{r}solution pump\end{tabular}}}}%
    \put(0.43835164,0.25550465){\makebox(0,0)[rt]{\lineheight{1.25}\smash{\begin{tabular}[t]{r}$\dot{m}^\r{v}_{\r{Ref,E,in}}$\end{tabular}}}}%
    \put(0.43835164,0.22272043){\makebox(0,0)[rt]{\lineheight{1.25}\smash{\begin{tabular}[t]{r}$\dot{m}^\r{l}_{\r{Ref,E,in}}$\end{tabular}}}}%
    \put(0.43835165,0.3696313){\makebox(0,0)[rt]{\lineheight{1.25}\smash{\begin{tabular}[t]{r}$\dot{m}_{\r{Ref,C,out}}$\end{tabular}}}}%
    \put(0.68200217,0.22473824){\color[rgb]{0,0,0}\makebox(0,0)[lt]{\lineheight{1.70454073}\smash{\begin{tabular}[t]{l}$\dot{m}_{\r{PSo,A,in}}$\end{tabular}}}}%
    \put(0.68200217,0.29627477){\color[rgb]{0,0,0}\makebox(0,0)[lt]{\lineheight{1.70454073}\smash{\begin{tabular}[t]{l}$\dot{m}_{\r{PSo,SHX,out}}$\end{tabular}}}}%
    \put(0.68200217,0.36991323){\color[rgb]{0,0,0}\makebox(0,0)[lt]{\lineheight{1.70454073}\smash{\begin{tabular}[t]{l}$\dot{m}_{\r{PSo,SHX,in}}$\end{tabular}}}}%
    \put(0.60367129,0.3696313){\makebox(0,0)[rt]{\lineheight{1.25}\smash{\begin{tabular}[t]{r}$\dot{m}_{\r{RSo,SHX,out}}$\end{tabular}}}}%
    \put(0.6036713,0.3018853){\makebox(0,0)[rt]{\lineheight{1.25}\smash{\begin{tabular}[t]{r}$\dot{m}_{\r{RSo,SHX,in}}$\end{tabular}}}}%
    \put(0.6036713,0.22272043){\makebox(0,0)[rt]{\lineheight{1.25}\smash{\begin{tabular}[t]{r}$\dot{m}_{\r{RSo,A,out}}$\end{tabular}}}}%
    \put(0.92886697,0.19571678){\color[rgb]{0,0,0}\makebox(0,0)[t]{\lineheight{1.70454276}\smash{\begin{tabular}[t]{c}cooling water \end{tabular}}}}%
    \put(0.9264246,0.17136318){\color[rgb]{0,0,0}\makebox(0,0)[t]{\lineheight{1.70454276}\smash{\begin{tabular}[t]{c}circuit\end{tabular}}}}%
    \put(0.2514415,0.03083835){\color[rgb]{0,0,0}\makebox(0,0)[lt]{\lineheight{1.70454276}\smash{\begin{tabular}[t]{l}evaporator (E)\end{tabular}}}}%
    \put(0.78745073,0.03083835){\color[rgb]{0,0,0}\makebox(0,0)[rt]{\lineheight{1.70454276}\smash{\begin{tabular}[t]{r}absorber (A)\end{tabular}}}}%
    \put(0.5360673,0.00513059){\color[rgb]{0,0,0}\makebox(0,0)[rt]{\lineheight{1.70454276}\smash{\begin{tabular}[t]{r}recirculation pump\\\end{tabular}}}}%
    \put(0.79309703,0.17275435){\color[rgb]{0,0,0}\makebox(0,0)[lt]{\lineheight{1.7045418}\smash{\begin{tabular}[t]{l}$\dot{m}_{\r{W,A,out}}$\end{tabular}}}}%
    \put(0.7950625,0.12254023){\color[rgb]{0,0,0}\makebox(0,0)[lt]{\lineheight{1.7045418}\smash{\begin{tabular}[t]{l}$\dot{m}_{\r{W,A,in}}$\end{tabular}}}}%
    \put(0.24633608,0.17351573){\makebox(0,0)[rt]{\lineheight{1.25}\smash{\begin{tabular}[t]{r}$\dot{m}_{\r{W,E,out}}$\end{tabular}}}}%
    \put(0.24633608,0.12436654){\makebox(0,0)[rt]{\lineheight{1.25}\smash{\begin{tabular}[t]{r}$\dot{m}_{\r{W,E,in}}$\end{tabular}}}}%
    \put(0,0){\includegraphics[width=\unitlength,page=5]{AHPS_modelling_V6b.pdf}}%
    \put(0.68376633,0.10218609){\color[rgb]{0,0,0}\makebox(0,0)[lt]{\lineheight{1.70454073}\smash{\begin{tabular}[t]{l}$\dot{m}_{\r{RSo,HX,A,out}}$\end{tabular}}}}%
    \put(0.35626786,0.11352023){\makebox(0,0)[rt]{\lineheight{1.25}\smash{\begin{tabular}[t]{r}$\dot{m}^\r{v}_{\r{Ref,HX,E,out}}$\end{tabular}}}}%
    \put(0.35626788,0.08368221){\makebox(0,0)[rt]{\lineheight{1.25}\smash{\begin{tabular}[t]{r}$\dot{m}^\r{l}_{\r{Ref,HX,E,out}}$\end{tabular}}}}%
    \put(0.35626788,0.18401538){\makebox(0,0)[rt]{\lineheight{1.25}\smash{\begin{tabular}[t]{r}$\dot{m}_{\r{Ref,rec}}$\end{tabular}}}}%
    \put(0.51987569,0.16467732){\makebox(0,0)[t]{\lineheight{1.25}\smash{\begin{tabular}[t]{c}$\dot{m}_{\r{Ref,GRl}}$\end{tabular}}}}%
    \put(0.12300241,0.1639461){\color[rgb]{0,0,0}\makebox(0,0)[t]{\lineheight{1.70454276}\smash{\begin{tabular}[t]{c}chilled water \end{tabular}}}}%
    \put(0.1205602,0.13959262){\color[rgb]{0,0,0}\makebox(0,0)[t]{\lineheight{1.70454276}\smash{\begin{tabular}[t]{c}circuit\end{tabular}}}}%
    \put(0.37593233,0.05873596){\color[rgb]{1,1,1}\makebox(0,0)[t]{\lineheight{1.25}\smash{\begin{tabular}[t]{c}$m_{\r{Ref,E}}$\end{tabular}}}}%
    \put(0.73357318,0.14739037){\color[rgb]{0,0,0}\makebox(0,0)[lt]{\lineheight{1.25}\smash{\begin{tabular}[t]{l}$\dot{Q}_{\r{A}}$\end{tabular}}}}%
    \put(0.30369747,0.14940171){\makebox(0,0)[rt]{\lineheight{1.25}\smash{\begin{tabular}[t]{r}$\dot{Q}_{\r{E}}$\end{tabular}}}}%
    \put(0.4355898,0.02826447){\color[rgb]{0,0,0}\makebox(0,0)[lt]{\lineheight{1.70454276}\smash{\begin{tabular}[t]{l}low pressure gas room (GRl)\end{tabular}}}}%
    \put(0,0){\includegraphics[width=\unitlength,page=6]{AHPS_modelling_V6b.pdf}}%
    \put(0.01540614,0.36229704){\color[rgb]{0,0,0}\makebox(0,0)[lt]{\lineheight{1.70454276}\smash{\begin{tabular}[t]{l}fluid legend:\end{tabular}}}}%
    \put(0.04429953,0.3358202){\color[rgb]{0,0,0}\makebox(0,0)[lt]{\lineheight{1.70454276}\smash{\begin{tabular}[t]{l}refrigerant (Ref)\end{tabular}}}}%
    \put(0,0){\includegraphics[width=\unitlength,page=7]{AHPS_modelling_V6b.pdf}}%
    \put(0.04429953,0.3093434){\color[rgb]{0,0,0}\makebox(0,0)[lt]{\lineheight{1.70454276}\smash{\begin{tabular}[t]{l}rich solution (RSo)\end{tabular}}}}%
    \put(0,0){\includegraphics[width=\unitlength,page=8]{AHPS_modelling_V6b.pdf}}%
    \put(0.04451053,0.2829088){\color[rgb]{0,0,0}\makebox(0,0)[lt]{\lineheight{1.70454276}\smash{\begin{tabular}[t]{l}poor solution (PSo)\end{tabular}}}}%
    \put(0,0){\includegraphics[width=\unitlength,page=9]{AHPS_modelling_V6b.pdf}}%
    \put(0.04429953,0.2564742){\color[rgb]{0,0,0}\makebox(0,0)[lt]{\lineheight{1.70454276}\smash{\begin{tabular}[t]{l}water (W) in external \end{tabular}}}}%
    \put(0.04429953,0.2320243){\color[rgb]{0,0,0}\makebox(0,0)[lt]{\lineheight{1.70454276}\smash{\begin{tabular}[t]{l}hydraulic circuits\end{tabular}}}}%
    \put(0,0){\includegraphics[width=\unitlength,page=10]{AHPS_modelling_V6b.pdf}}%
    \put(0.38491449,0.40294233){\color[rgb]{1,1,1}\makebox(0,0)[t]{\lineheight{1.25}\smash{\begin{tabular}[t]{c}$m_{\r{Ref,C}}$\end{tabular}}}}%
  \end{picture}%
\endgroup%

%% file: pics/SHX_dyn.pdf_tex
\begingroup%
  \makeatletter%
  \providecommand\color[2][]{%
    \errmessage{(Inkscape) Color is used for the text in Inkscape, but the package 'color.sty' is not loaded}%
    \renewcommand\color[2][]{}%
  }%
  \providecommand\transparent[1]{%
    \errmessage{(Inkscape) Transparency is used (non-zero) for the text in Inkscape, but the package 'transparent.sty' is not loaded}%
    \renewcommand\transparent[1]{}%
  }%
  \providecommand\rotatebox[2]{#2}%
  \newcommand*\fsize{\dimexpr\f@size pt\relax}%
  \newcommand*\lineheight[1]{\fontsize{\fsize}{#1\fsize}\selectfont}%
  \ifx\svgwidth\undefined%
    \setlength{\unitlength}{269.43874569bp}%
    \ifx\svgscale\undefined%
      \relax%
    \else%
      \setlength{\unitlength}{\unitlength * \real{\svgscale}}%
    \fi%
  \else%
    \setlength{\unitlength}{\svgwidth}%
  \fi%
  \global\let\svgwidth\undefined%
  \global\let\svgscale\undefined%
  \makeatother%
  \begin{picture}(1,0.0925809)%
    \lineheight{1}%
    \setlength\tabcolsep{0pt}%
    \put(0,0){\includegraphics[width=\unitlength,page=1]{SHX_dyn.pdf}}%
    \put(0.09155096,0.04596519){\makebox(0,0)[t]{\lineheight{1.25}\smash{\begin{tabular}[t]{c}$T_{\r{RSo,SHX,out,ss}}$\end{tabular}}}}%
    \put(0,0){\includegraphics[width=\unitlength,page=2]{SHX_dyn.pdf}}%
    \put(0.52607169,0.02922168){\makebox(0,0)[t]{\lineheight{1.25}\smash{\begin{tabular}[t]{c}$\tau=\frac{m_{\r{RSo,SHX}}}{\dot{m}_{\r{RSo}}}$\end{tabular}}}}%
    \put(0.35382742,0.04596519){\makebox(0,0)[t]{\lineheight{1.25}\smash{\begin{tabular}[t]{c}$h_{\r{RSo,SHX,out,ss}}$\end{tabular}}}}%
    \put(0,0){\includegraphics[width=\unitlength,page=3]{SHX_dyn.pdf}}%
    \put(0.68655597,0.04596519){\makebox(0,0)[t]{\lineheight{1.25}\smash{\begin{tabular}[t]{c}$h_{\r{RSo,SHX,out}}$\end{tabular}}}}%
    \put(0,0){\includegraphics[width=\unitlength,page=4]{SHX_dyn.pdf}}%
    \put(0.92366461,0.04596519){\makebox(0,0)[t]{\lineheight{1.25}\smash{\begin{tabular}[t]{c}$T_{\r{RSo,SHX,out}}$\end{tabular}}}}%
    \put(0,0){\includegraphics[width=\unitlength,page=5]{SHX_dyn.pdf}}%
    \put(0.8072063,0.02376532){\makebox(0,0)[t]{\lineheight{1.25}\smash{\begin{tabular}[t]{c}\scriptsize{\autoref{eq:h_LiBrH2O}}\end{tabular}}}}%
    \put(0,0){\includegraphics[width=\unitlength,page=6]{SHX_dyn.pdf}}%
    \put(0.22254599,0.02376532){\makebox(0,0)[t]{\lineheight{1.25}\smash{\begin{tabular}[t]{c}\scriptsize{\autoref{eq:h_LiBrH2O}}\end{tabular}}}}%
    \put(0.52819286,0.07702272){\makebox(0,0)[t]{\lineheight{1.25}\smash{\begin{tabular}[t]{c}first-order delay\end{tabular}}}}%
  \end{picture}%
\endgroup%